\documentclass[prc,preprint,
  showpacs,showkeys,lengthcheck,
  nofootinbib,tightenlines,onecolumn,notitlepage,
  preprintnumbers,superscriptaddress]{revtex4-1}

\usepackage{graphicx}
\usepackage[usenames,dvipsnames]{xcolor}
\graphicspath{{figures/}{figures/diagrams/}}

\usepackage{xr-hyper}
\usepackage{hyperref}
\hypersetup{colorlinks=true,citecolor=blue,linkcolor=blue,urlcolor=blue}
\usepackage{diagbox}
\usepackage{booktabs}

\usepackage{amsmath,amssymb,amsfonts}
\usepackage{mathrsfs}
\usepackage{slashed}
\usepackage{bm}

\usepackage[final]{changes} 
\definechangesauthor[color=orange!76!black]{Adam}
\definechangesauthor[color=green!50!black]{Wim}
\definechangesauthor[color=blue]{rev}  

\usepackage{outlines}
\usepackage{enumitem}
\setenumerate[1]{label=\Roman*.}
\setenumerate[2]{label=\Alph*.}
\setenumerate[3]{label=\roman*.}
\setenumerate[4]{label=\alph*.}

\newcommand{\LDR}{\ensuremath{\overline{\mu}}}
\newcommand{\companion}{the companion paper~\cite{Freese:2022yur}}

\begin{document}

\title{Spatial densities of the photon on the light front}

\author{Adam Freese}
\email{afreese@uw.edu}
\affiliation{Department of Physics, University of Washington, Seattle, WA 98195, USA}

\author{Wim Cosyn}
\email{wcosyn@fiu.edu}
\affiliation{Department of Physics, Florida International University, Miami, FL 33199, USA}
\affiliation{Department of Physics and Astronomy, Ghent University, B9000 Gent, Belgium}

\begin{abstract}
  The light front densities of momentum, angular momentum, and intrinsic pressure
  are calculated for the photon,
  both in the free case and at leading order in quantum electrodynamics.
  In the latter case, we moreover decompose the form factors into
  photon and electron contributions.
  Circularly and linearly polarized photons are both considered,
  with the latter containing significant azimuthal modulations
  in both the momentum density and in intrinsic stresses.
  We find that the D-term of the photon is positive instead of negative,
  and accordingly the intrinsic radial pressure of the photon is negative.
  Despite this, the radiation pressure exerted by the photon is positive.
  We illustrate through explicit calculation how the intrinsic pressure
  associated with the D-term and the radiation pressure exerted by
  the photon are different quantities.
\end{abstract}

\preprint{NT@UW-2207}

\maketitle


\section{Introduction}
\label{sec:intro}

The energy-momentum tensor (EMT) has become an increasingly relevant
topic in the hadron physics community,
owing largely to its implications for the proton mass
puzzle~\cite{Ji:1994av,Ji:1995sv,Lorce:2017xzd,Hatta:2018sqd,Metz:2020vxd,Ji:2021mtz,Lorce:2021xku}
and the proton spin
puzzle~\cite{Ashman:1987hv,Ji:1996ek,Leader:2013jra}.
Matrix elements of the EMT between hadron states encode a slate of
gravitational form factors (GFFs),
which are the primary objects of study in works on the subject.
In addition to the puzzles pertaining to properties of the proton,
much attention has been given to studies of
mechanical properties~\cite{Polyakov:2002yz,Perevalova:2016dln,Polyakov:2018zvc}
and spatial densities~\cite{Polyakov:2018zvc,Lorce:2018egm,Lorce:2020onh,Freese:2021czn}
related to the GFFs, especially the so-called
D-term~\cite{Polyakov:1999gs} or Druck term~\cite{Polyakov:2018rew}.

In \companion,
we determined the two-dimensional light front densities associated
with the GFFs of spin-one targets.
The formalism presented there was focused primary on massive spin-one targets,
but is applicable (with little modification) to massless targets as well.
The photon is an especially pertinent case of a spin-one target,
which has previously been a considered as target in theoretical hadron physics
studies~\cite{Witten:1977ju,Uematsu:1980qy,Friot:2006mm,ElBeiyad:2008ss,Gabdrakhmanov:2012aa}.
Its gravitational form factors have also been calculated
long ago~\cite{Berends:1975ah,Milton:1977je} to leading order in quantum electrodynamics (QED).
More recently, the QED electron at leading order has been used as a toy model
for studying the proton mass puzzle~\cite{Rodini:2020pis},
giving precedent for using dressed particles from QED as a playground
for better understanding issues related to the EMT for hadron physics.
It is therefore worthwhile considering what the EMT densities for
a photon target are---both for a free photon,
and for the QED photon at leading order.

In fact, the photon has a special peculiarity that makes it especially pertinent.
It is widely believed that negativity of the D-term is a necessary condition
for stability of a target.
In particular, it is believed to be related to positivity of the radial
pressure~\cite{Polyakov:2018zvc,Lorce:2018egm,Freese:2021czn}.
The negativity condition already been called into question
by the D-term of the QED electron being infinite and positive~\cite{Metz:2021lqv},
but controversy remains on whether this is a pathology of long-range forces
that can be eliminated by a redefinition~\cite{Varma:2020crx}.
We find, however, that the D-term of the photon is positive
even for a free photon.
\replaced{
It is worth noting, however, that the photon being massless prevents it
from decaying, so its exception to the postulated stability criterion
may not be an issue.
}{
further eroding the negativity condition
and also raising questions about the meaning of the ``pressure''
encoded by the GFFs.
}

We initiate this study alongside \companion,
precisely because the formalism of light front densities of
spin-one targets is necessary to even consider the question.
Although the physical meaning of Breit frame densities remains
controversial~\cite{Fleming:1974af,Burkardt:2002hr,Miller:2018ybm,Lorce:2018egm,Jaffe:2020ebz,Lorce:2020onh,Freese:2021czn,Epelbaum:2022fjc},
it is sometimes claimed that they give information about densities
in the rest frame of the target, possibly subject to
``relativistic corrections''~\cite{Polyakov:2018zvc}.
It is clear---even if Breit frame densities do have such a physical
meaning---that \added{the meaning ascribed by} such a framework is completely inapplicable to a photon target.
On the other hand, the physical meaningfulness of light front densities
requires only the possibility of localizing the target in the transverse
plane~\cite{Burkardt:2002hr,Miller:2018ybm,Freese:2021czn}---a feat
that can, in principle, be accomplished for a photon just as much
as for any massive target.

This work is organized as follows.
Sec.~\ref{sec:formalism} presents the few changes to the formalism of
\companion~needed to investigate the massless photon as a target.
Sec.~\ref{sec:free} then obtains the densities of a free photon,
and Sec.~\ref{sec:QED} subsequently obtains the densities of the QED
photon at leading order.
Sec.~\ref{sec:rad} investigates the light front density of
transverse pressure exerted \emph{by} a photon, i.e., radiation pressure,
in order to highlight the difference between this and
the \emph{intrinsic} pressure that is encoded by the D-term.


\section{Formalism}
\label{sec:formalism}

The majority of the necessary formalism is presented in \companion.
However, the photon possesses a few special properties as a massless gauge boson.
Firstly, there are fewer independent gravitational form factors than the
massive case---three instead of six.
Secondly, there are no transversely polarized states,
but superpositions of helicity states instead give elliptically or linearly
polarized states.
This section will address these special properties of the photon.


\subsection{Gravitational form factors of a massless photon}

The photon has fewer gravitational form factors than other spin-one targets
by virtue of gauge invariance.
This can be seen by considering the massive spin-one EMT in
Eq.~(\ref*{P1:eqn:emt:massive}) of \companion (see also
Refs.~\cite{Holstein:2006ge,Abidin:2008ku,Taneja:2011sy,Cosyn:2019aio,Freese:2019bhb,Polyakov:2019lbq}):
\begin{align}
  \label{eqn:emt:massive}
  \langle p'\lambda' | T^{\mu\nu}(0) | p \lambda \rangle
  &=
  2 P^\mu P^\nu
  \left[
    -(\varepsilon\cdot\varepsilon'^*)
    \mathcal{G}_1(t)
    + \frac{(\varepsilon\cdot\Delta)(\varepsilon'^*\cdot\Delta)}{2M^2}
    \mathcal{G}_2(t)
    \right]
  \notag \\ &
  +
  \frac{\Delta^\mu\Delta^\nu-\Delta^2g^{\mu\nu}}{2}
  \left[
    -(\varepsilon\cdot\varepsilon'^*)
    \mathcal{G}_3(t)
    + \frac{(\varepsilon\cdot\Delta)(\varepsilon'^*\cdot\Delta)}{2M^2}
    \mathcal{G}_4(t)
    \right]
  +
  \frac{1}{2}
  P^{\{\mu}
  \Big[
    \varepsilon'^{*\nu\}} (\varepsilon\cdot\Delta)
    -
    \varepsilon^{\nu\}} (\varepsilon'^*\cdot\Delta)
    \Big]
  \mathcal{G}_5(t)
  \notag \\ &
  +
  \frac{1}{4}
  \Big[
    \Delta^{\{\mu}
    \Big(
      \varepsilon'^{*\nu\}} (\varepsilon\cdot\Delta)
      +
      \varepsilon^{\nu\}} (\varepsilon'^*\cdot\Delta)
      \Big)
    -
    \varepsilon^{\{\mu}\varepsilon'^{*\nu\}} \Delta^2
    -
    2 g^{\mu\nu}
    (\varepsilon\cdot\Delta)(\varepsilon'^*\cdot\Delta)
    \Big]
  \mathcal{G}_6(t)
  \,,
\end{align}
but with $M$ signifying an arbitrary positive quantity with units of energy
instead of the photon mass (which is of course zero).
Gauge invariance requires that the EMT should be invariant
under the substitutions:
\begin{subequations}
  \begin{align}
    \varepsilon(p)
    &\mapsto
    \varepsilon(p)
    -
    i\tilde\chi(p) p
    \\
    \varepsilon'^*(p')
    &\mapsto
    \varepsilon'^*(p')
    +
    i\tilde\chi(p') p'
    \,,
  \end{align}
\end{subequations}
which in turn places strong constraints on the form factors
appearing in Eq.~(\ref{eqn:emt:massive}).
These constraints can be used to remove any dependence
of the EMT matrix element on the arbitrary constant $M$,
which now depends on only three independent form factors.

There are multiple bases that can be used to represent
the three independent form factors.
For the sake of continuity of the literature,
we use the notation of Milton~\cite{Milton:1977je} in particular,
in which the EMT of the photon is decomposed as:
\begin{subequations}
  \label{eqn:Milton}
  \begin{align}
    \langle p' \lambda' | T^{\mu\nu}(0) | p \lambda \rangle
    &=
    \Theta_1^{\mu\nu} F_1(t)
    +
    \Theta_2^{\mu\nu} F_2(t)
    +
    \Theta_3^{\mu\nu} F_3(t)
    \\
    \Theta_1^{\mu\nu}
    &=
    -2 P^\mu P^\nu
    (\varepsilon\cdot\varepsilon'^*)
    +
    \frac{1}{2}\Big(\Delta^\mu\Delta^\nu-\Delta^2g^{\mu\nu}\Big)
    (\varepsilon\cdot\varepsilon'^*)
    +
    P^{\{\mu} \Big[
      \varepsilon'^{*\nu\}} (\Delta\cdot\varepsilon)
      -
      \varepsilon^{\nu\}} (\Delta\cdot\varepsilon^*)
      \Big]
    \notag \\ &
    -
    \frac{1}{2} \Big[
      \Delta^{\{\mu}\big(
      \varepsilon'^{*\nu\}} (\Delta\cdot\varepsilon)
      +
      \varepsilon^{\nu\}} (\Delta\cdot\varepsilon'^*)
      \big)
      - \varepsilon^{\{\mu}\varepsilon'^{*\nu\}} \Delta^2
      - 2g^{\mu\nu} (\Delta\cdot\varepsilon)(\Delta\cdot\varepsilon'^*)
      \Big]
    \\
    \Theta_2^{\mu\nu}
    &=
    2
    \Big(
      2(\Delta\cdot\varepsilon)(\Delta\cdot\varepsilon'^*)
      - \Delta^2 (\varepsilon\cdot\varepsilon'^*)
      \Big)
    (\Delta^\mu \Delta^\nu - \Delta^2 g^{\mu\nu})
    \\
    \Theta_3^{\mu\nu}
    &=
    4
    P^\mu P^\nu
    \Big(
      2(\Delta\cdot\varepsilon)(\Delta\cdot\varepsilon'^*)
      - \Delta^2 (\varepsilon\cdot\varepsilon'^*)
      \Big)
    \,,
  \end{align}
\end{subequations}
where the brackets signify symmetrization:
$a^{\{\mu}b^{\nu\}} = a^\mu b^\nu + a^\nu b^\mu$.
It is worth noting that $\Theta_1^{\mu\nu}$ is a traceless tensor,
whereas $\Theta_{2,3}^{\mu\nu}$ are not traceless.
However, all three structures contain a non-zero
traceless piece, i.e., a piece that transforms
under the $(1,1)$ representation of the Lorentz group.

\added{
A peculiarity of Milton's form factor breakdown
is that $F_2(t)$ and $F_3(t)$ are unitful, with units GeV$^{-2}$.
Normally, one might introduce a factor of the target mass $M^{-2}$
into the tensors $\Theta_{2,3}^{\mu\nu}$ to keep the form factors unitless,
but this cannot be done for a massless target.
Analyticity also prevents a pole at $t=0$ from appearing,
so the only unit GeV$^{-2}$ quantities that could factor out of $F_{2,3}(t)$
are masses of other particles that interact with the photon.
This of course means that $F_{2,3}(t)$ must vanish for a free photon.
It also means that a scale is introduced explicitly by the interaction
with massive charged particles.
It's a choice of convention to leave $F_{2,3}(t)$ unitful
instead of introducing a factor of $m_e^{-2}$ into the tensors
$\Theta_{2,3}^{\mu\nu}$.
}

In \companion, an alternative slate of \emph{effective}
form factors was proposed in its Eq.~(\ref*{P1:eqn:emt:spin1}).
\added{
Restated here for convenience, the expression is:
}
\begin{align}
  \label{eqn:emt:spin1}
  \langle p'\lambda' | T^{\mu\nu}(0) | p \lambda \rangle
  \bigg|_{\Delta^+=0}
  &=
  2 P^\mu P^\nu
  \mathcal{A}_{\lambda'\lambda}(\boldsymbol{\Delta}_\perp)
  -
  i
  \frac{ P^{\{\mu}\epsilon^{\nu\}P\Delta n} }{ (P\cdot n) }
  \mathcal{J}_{\lambda'\lambda}(\boldsymbol{\Delta}_\perp)
  +
  \frac{ \Delta^\mu \Delta^\nu - \Delta^2 g^{\mu\nu} }{2}
  \mathcal{D}_{\lambda'\lambda}(\boldsymbol{\Delta}_\perp)
  \notag
  \\ &
  +
  \frac{P^{\{\mu}n^{\nu\}}}{(P\cdot n)}
  \mathcal{E}_{\lambda'\lambda}(\boldsymbol{\Delta}_\perp)
  +
  \frac{n^\mu n^\nu}{(P\cdot n)^2}
  \mathcal{H}_{\lambda'\lambda}(\boldsymbol{\Delta}_\perp)
  +
  i
  \frac{ n^{\{\mu}\epsilon^{\nu\}P\Delta n} }{ (P\cdot n)^2 }
  \mathcal{K}_{\lambda'\lambda}(\boldsymbol{\Delta}_\perp)
  \,.
\end{align}
\replaced{
These ``effective form factors'' depend on helicity and
on the four-vector $n$ defining the light front,
and are thus really light front helicity amplitudes rather than
proper form factors.
We consider them to be ``effective form factors''
because their 2D Fourier transforms conveniently
}{
These effective form factors depend on helicity
in addition to momentum transfer,
but several of these have a more direct physical interpretation,
in that their light front Fourier transforms
}
give conventional Galilean densities.
\replaced{
The helicity amplitudes for the fixed-helicity case,
which function as effective form factors for helicity states,
can be expressed
}
{
For photons in light front helicity states in particular,
these effective form factors can be expressed
}
in terms of the Milton form factors as:
\begin{subequations}
  \label{eqn:gff:helicity}
  \begin{align}
    \mathcal{A}_{\lambda\lambda}(t) &= F_1(t)
    \\
    \mathcal{J}_{\lambda\lambda}(t) &= \lambda F_1(t)
    \\
    \mathcal{D}_{\lambda\lambda}(t) &= F_1(t)
    \\
    \mathcal{E}_{\lambda\lambda}(t) &= t F_1(t)
    \\
    \mathcal{H}_{\lambda\lambda}(t) &= \frac{t^2}{4} F_1(t)
    \\
    \mathcal{K}_{\lambda\lambda}(t) &= - \lambda \frac{t}{4} F_1(t)
    \,.
  \end{align}
\end{subequations}
\replaced{
Light front densities
}{
The effective form factors
}
for helicity states thus depend only on
a single independent form factor $F_1(t)$.
The other Milton form factors appear instead in the
\replaced{
helicity-flip amplitudes:
}{
effective form factors for helicity-flip matrix elements:
}
\begin{subequations}
  \label{eqn:gff:flip}
  \begin{align}
    \mathcal{A}_{\pm\mp}(t) &= 2 t F_3(t)
    \\
    \mathcal{J}_{\pm\mp}(t) &= 0
    \\
    \mathcal{D}_{\pm\mp}(t) &= 2 t F_2(t)
    \\
    \mathcal{E}_{\pm\mp}(t) &= 0
    \\
    \mathcal{H}_{\pm\mp}(t) &= 0
    \\
    \mathcal{K}_{\pm\mp}(t) &= 0
    \,.
  \end{align}
\end{subequations}
These two form factors thus contribute only to the densities of
photons with non-circular polarizations.


\subsection{Linear polarization}

We consider linearly polarized photons as an extreme case of photons
with a non-circular polarization
in order to study the effects of $F_2(t)$ and $F_3(t)$ on photon densities.
Horizontal polarization (electric fields oscillating along the $x$ axis)
are considered for definiteness.
In light of the helicity vector conventions used in this work
(which are explicitly given in Eq.~(\ref*{P1:eqn:helicity:pm}) of \companion),
the horizontal and vertical polarization vectors can be written:
\begin{subequations}
  \begin{align}
    \varepsilon_H
    &=
    \frac{ -\varepsilon_+ + \varepsilon_- }{\sqrt{2}}
    \\
    \varepsilon_V
    &=
    \frac{ i\varepsilon_+ + i\varepsilon_- }{\sqrt{2}}
    \,.
  \end{align}
\end{subequations}
Accordingly, helicity-flip contributions are present in the effective
form factors for linearly polarized photons.
In particular:
\begin{subequations}
  \label{eqn:gffs:linear}
  \begin{align}
    \mathcal{A}_{\mathrm{lin}}(\boldsymbol{\Delta}_\perp)
    &=
    F_1(t) - 2t F_3(t) \cos2\phi
    \\
    \mathcal{J}_{\mathrm{lin}}(\boldsymbol{\Delta}_\perp)
    &=
    0
    \\
    \mathcal{D}_{\mathrm{lin}}(\boldsymbol{\Delta}_\perp)
    &=
    F_1(t) - 2t F_2(t) \cos2\phi
    \\
    \mathcal{E}_{\mathrm{lin}}(\boldsymbol{\Delta}_\perp)
    &=
    t F_1(t)
    \\
    \mathcal{H}_{\mathrm{lin}}(\boldsymbol{\Delta}_\perp)
    &=
    \frac{t^2}{4} F_1(t)
    \\
    \mathcal{K}_{\mathrm{lin}}(\boldsymbol{\Delta}_\perp)
    &=
    0
    \,,
  \end{align}
\end{subequations}
where $\phi$ is the angle between the direction of the transverse electric field
and the direction of $\boldsymbol{\Delta}_\perp$.
Consequently, linearly polarized photons can potentially have
$\cos2\phi$ modulations in their densities.


\section{EMT and densities of a free photon}
\label{sec:free}

The energy-momentum tensor of the free photon field
is~\cite{Bessel:1921emt,Jackson:1998nia}:
\begin{align}
  \label{eqn:EMT:photon}
  T^{\mu\nu}(x)
  =
  F^{\mu\lambda} F_{\lambda}^{\phantom{\lambda}\nu}
  +
  \frac{1}{4} g^{\mu\nu} F^{\lambda\rho} F_{\lambda\rho}
  \,.
\end{align}
To find matrix elements of this EMT between single-photon states,
a normal mode decomposition of the four-potential can be used:
\begin{align}
  A^\mu(x)
  =
  \sum_\lambda
  \int \frac{\mathrm{d}k^+ \mathrm{d}^2\mathbf{k}_\perp}{2k^+(2\pi)^3}
  \Big\{
    \varepsilon^\mu(k,\lambda)
    a(k,\lambda)
    e^{-ik\cdot x}
    +
    \varepsilon^{*\mu}(k,\lambda)
    a^\dagger(k,\lambda)
    e^{+ik\cdot x}
    \Big\}
  \,.
\end{align}
Using normal ordering with Eq.~(\ref{eqn:EMT:photon})
(to avoid an infinite vacuum contribution),
we find:
\begin{align}
  \label{eqn:photon:free}
  \langle p' \lambda' | \mathop{:}\!T^{\mu\nu}(0)\!\mathop{:} | p \lambda \rangle
  =
  \Theta_1^{\mu\nu}
  \,,
\end{align}
where $\Theta_1^{\mu\nu}$ is as defined in Eq.~(\ref{eqn:Milton}).
Consequently, for the free photon, one has:
\begin{subequations}
  \begin{align}
    & F_1^{\mathrm{free}}(t) = 1
    \\
    & F_2^{\mathrm{free}}(t) = F_3^{\mathrm{free}}(t) = 0
    \,.
  \end{align}
\end{subequations}
Thus, linearly polarized free photons do not have
angular modulations in their densities.
\added{
Since $F_2(t)$ and $F_3(t)$ are unitful form factors,
and a free photon does not have an obvious mass scale,
these quantities must necessarily vanish in the free case.
}

\added{
We remind the reader that of the six light front helicity amplitudes
(or effective form factors),
only $\mathcal{A}$, $\mathcal{J}$ and $\mathcal{D}$
contribute to the Galilean densities.
}
All free photons have the following effective form factors:
\begin{align}
  \mathcal{A}^{\mathrm{free}}(t)
  =
  \mathcal{D}^{\mathrm{free}}(t)
  =
  1
  \,,
\end{align}
and helicity states in particular have:
\begin{align}
  \mathcal{J}^{\mathrm{free}}_{\lambda}(t)
  =
  \lambda
  \,.
\end{align}
Since the free photon is pointlike, these results are almost trivial.
The momentum and spin sum rules
$\mathcal{A}(0)=1$ and $\mathcal{J}_\lambda(0)=\lambda$ are satisfied,
and the corresponding $P^+$ and angular momentum densities are
delta functions at the origin.
The $D$-term has the surprising property, however,
that $\mathcal{D}(0)=1 > 0$.
It has widely been postulated---first in Ref.~\cite{Perevalova:2016dln}
and later elsewhere, see
Refs.~\cite{Polyakov:2018zvc,Lorce:2018egm,Freese:2021czn}
for instance---that $\mathcal{D}(0) < 0$ is required for stability of a system.
However, the free photon clearly violates this condition.
\replaced{
It is unnecessary for massless states to satisfy the stability criterion,
however, since their masslessness prevents them from decay.
}{
This suggests that negativity of $\mathcal{D}(0)$ cannot be taken as a universal
necessary condition for stability after all.
}

In fact, the result $\mathcal{D}(0)=1$ is true for the photon by virtue
of gauge invariance, and will also hold when QED corrections are considered.
This can be seen immediately from
Eqs.~(\ref{eqn:gff:helicity}) and (\ref{eqn:gff:flip}).
One must have $F_1(0) = 1$ to satisfy the sum rule $\mathcal{A}(0)=1$,
but as a consequence of gauge invariance,
we also have $\mathcal{D}(0) = F_1(0)$.
Thus, the introduction of interactions will not make $\mathcal{D}(0)$ negative.


\subsection{Densities and mechanical properties of the free photon}

Using Eq.~(\ref*{P1:eqn:p+:hel}) of \companion:
\begin{align}
  \label{eqn:p+:hel}
  \rho_{p^+}^{(\lambda)}(b_\perp)
  &=
  P^+
  \int \frac{\mathrm{d}^2\boldsymbol{\Delta}_\perp}{(2\pi)^2}
  \mathcal{A}_{\lambda\lambda}(t)
  e^{-i\boldsymbol{\Delta}_\perp\cdot\mathbf{b}_\perp}
  \,,
\end{align}
the $P^+$ density of the free photon is:
\begin{align}
  \rho_{P^+}(\mathbf{b}_\perp)
  =
  P^+
  \delta^{(2)}(\mathbf{b}_\perp)
  \,,
\end{align}
as expected of a point particle.
For helicity states, the angular momentum is similarly given by
a delta function at the transverse origin.

Obtaining the light front comoving stress tensor is more subtle.
From its definition in
Eq.~(\ref*{P1:eqn:Dtilde}) of \companion:
\begin{align}
  \label{eqn:Dtilde}
  \widetilde{\mathcal{D}}_\lambda(b_\perp)
  &=
  \frac{1}{4P^+}
  \int \frac{\mathrm{d}^2\boldsymbol{\Delta}_\perp}{(2\pi)^2}
  \mathcal{D}_{\lambda\lambda}(t)
  e^{-i\boldsymbol{\Delta}_\perp\cdot\mathbf{b}_\perp}
  \,,
\end{align}
the light front Polyakov stress potential of the free photon
can be written:
\begin{align}
  \widetilde{\mathcal{D}}_\gamma(b_\perp)
  =
  \frac{1}{4P^+}
  \delta^{(2)}(\mathbf{b}_\perp)
  =
  \frac{1}{4P^+}
  \lim_{\sigma\rightarrow0}
  \left\{
    \frac{1}{2\pi \sigma^2}
    e^{-\frac{1}{2\sigma^2}\mathbf{b}_\perp^2}
    \right\}
  \,.
\end{align}
It is prudent to use a Gaussian representation for the delta function,
especially since the Gaussian wave packet is more
physical than the fully localized state
(with only the former having a representative
in Hilbert space~\cite{vonNeumann:1955book}).
As discussed in Refs.~\cite{Miller:2018ybm,Freese:2021czn},
one should in general defer taking the $\sigma\rightarrow0$ limit
until after all other calculations have been performed,
except in cases where the dominated convergence theorem allows
this limit to be commuted inside any integrands or derivatives.
Dealing with Gaussian wave packets in this case specifically
allows the pathologies of delta functions to be avoided.

The stress tensor can be fully parametrized by its eigenpressures.
For the free photon (as well as for helicity states of interacting photons),
these are the radial and tangential pressures given
by~\cite{Lorce:2018egm,Freese:2021czn,Freese:2022yur}:
\deleted{Eqs.~(\ref*{P1:eqn:prt:hel}) of \companion:}
\begin{subequations}
  \label{eqn:prt:hel}
  \begin{align}
    \label{eqn:pr:hel}
    p_r^{(\lambda)}(b_\perp)
    &=
    p^{(\lambda)}(b_\perp)
    +
    \frac{s^{(\lambda)}(b_\perp)}{2}
    =
    \frac{1}{b_\perp}
    \frac{\mathrm{d} \widetilde{\mathcal{D}}_\lambda(b_\perp)}{\mathrm{d}b_\perp}
    \\
    p_t^{(\lambda)}(b_\perp)
    &=
    p^{(\lambda)}(b_\perp)
    -
    \frac{s^{(\lambda)}(b_\perp)}{2}
    =
    \frac{\mathrm{d}^2\widetilde{\mathcal{D}}_\lambda(b_\perp)}{\mathrm{d}b_\perp^2}
    \,.
  \end{align}
\end{subequations}
For the free photon wave packet in particular:
\begin{subequations}
  \begin{align}
    p_r(b_\perp;\sigma)
    &=
    -
    \frac{1}{4P^+}
    \frac{1}{2\pi \sigma^4}
    e^{-\frac{1}{2\sigma^2}\mathbf{b}_\perp^2}
    \\
    p_t(b_\perp;\sigma)
    &=
    -
    \frac{1}{4P^+}
    \frac{1}{2\pi \sigma^4}
    \left(1 - \frac{\mathbf{b}_\perp^2}{\sigma^2}\right)
    e^{-\frac{1}{2\sigma^2}\mathbf{b}_\perp^2}
    \,.
  \end{align}
\end{subequations}
The sign behavior of this is exactly the opposite of what is typically
expected for the radial and tangential pressures:
the radial pressure is negative-definite,
while the tangential pressure is negative at short distances from the origin
and positive at larger distances.
This is of course related to $\mathcal{D}(0)$ being positive
instead of negative.

Typically, the radial pressure in particular is used to define
static mechanical properties of a system
out of the expectation that it is positive-definite.
Despite the photon radial pressure being negative,
we can still proceed with the usual definitions.
The integral of the radial pressure is negative,
and diverges in the limit of wave packet localization:
\begin{align}
  \label{eqn:notzero}
  \int \mathrm{d}^2\mathbf{b}_\perp \,
  p_{r,\gamma}(b_\perp;\sigma)
  =
  -
  \frac{1}{4P^+}
  \frac{1}{\sigma^2}
  \,.
\end{align}
The $\mathbf{b}_\perp^2$-weighted moment is also negative, but finite:
\begin{align}
  \int \mathrm{d}^2\mathbf{b}_\perp \,
  \mathbf{b}_\perp^2
  p_{r,\gamma}(b_\perp;\sigma)
  =
  -
  \frac{1}{2P^+}
  <
  0
  \,.
\end{align}
Consequently, the mean squared mechanical radius,
as given by~\cite{Freese:2021czn,Panteleeva:2021iip}:
\begin{align}
  \label{eqn:r:mech}
  \langle b_\perp^2 \rangle_{\text{mech}}(\sigma)
  =
  \frac{
    \int \mathrm{d}^2\mathbf{b}_\perp \,
    \mathbf{b}_\perp^2
    p_{r,\gamma}(b_\perp;\sigma)
  }{
    \int \mathrm{d}^2\mathbf{b}_\perp \,
    p_{r,\gamma}(b_\perp;\sigma)
  }
  =
  2 \sigma^2
  \,,
\end{align}
is positive at $\sigma > 0$,
and goes to zero as $\sigma\rightarrow0$.
The mechanical radius of the free photon is zero,
as one would expect for a pointlike particle.


It appears almost absurd that the radial pressure of the free photon
should be negative.
After all, photons are known to exert positive
radiation pressure~\cite{Jackson:1998nia}.
Our finding does not contradict this known result, however.
As previously explained in Ref.~\cite{Freese:2021czn},
the stress tensor $T^{ij}$ as a whole consists of two pieces:
a piece that encodes the \emph{flow} of the target
\replaced{
in the transverse plane
}{
as a whole
}
(both due to average motion and wave function dispersion),
and a piece that encodes the stresses that would be measured by an observer
comoving with that \added{transverse} flow.
\replaced{
Previous works have not considered the contributions of hadron flow explicitly,
since their goal was to only describe intrinsic structure.
These works thus effectively studied only the pressure as seen by the comoving observer.
}{
Much of the earlier literature on EMT densities in hadrons
neglects the effects of hadron flow,
thus effectively studying only the pressure as seen by the comoving observer.
}
This quantity is identified as an \emph{intrinsic} pressure of the system,
and for the photon, this is negative.
\replaced{
The radiation pressure, however, includes the flow that has been so far not included explicitly.
}{
The radiation pressure, however, includes the flow that has been so far neglected.
}
We shall calculate transverse radial radiation pressure in
Sec.~\ref{sec:rad} and comment further there.


\section{Leading-order QED corrections}
\label{sec:QED}

In this section, we consider leading-order corrections to the photon's
gravitational form factors arising from quantum electrodynamics (QED).
The full QED Lagrangian can be written in the
Gupta-Bleuler formalism~\cite{Gupta:1949rh,Bleuler:1950cy} as:
\begin{align}
  \label{eqn:lagrangian:qed}
  \mathscr{L}
  =
  \bar\psi_0
  \left(
  \frac{i}{2}
  \overleftrightarrow{\slashed{\partial}}
  -
  e_0 \slashed{A}_0
  -
  m_0
  \right)
  \psi_0
  -
  \frac{1}{4} F_0^2
  -
  \frac{\lambda_0}{2} (\partial\cdot A_0)^2
  \,,
\end{align}
where we have explicitly noted that this expression
is in terms of the \emph{bare} (unrenormalized) fields.
The EMT for this Lagrangian is is~\cite{Embacher:1986pt,Freese:2021jqs}:
\begin{align}
  \label{eqn:emt:qed}
  T^{\mu\nu}
  =
  \frac{i}{4} \bar\psi_0
  \gamma^{\{\mu} \overleftrightarrow{\partial}^{\nu\}}
  \psi
  - \frac{1}{2} e_0 \bar{\psi}_0 \gamma^{\{\mu}_{\phantom{0}}A^{\nu\}}_0 \psi_0
  + F_0^{\phantom{0}\mu\sigma} F_{0\sigma}^{\phantom{0\sigma}\nu}
  - \lambda_0 (\partial\cdot A_0) \partial^{\{\mu}_{\phantom{0}} A_0^{\nu\}}
  -
  g^{\mu\nu} \mathscr{L}
  \,.
\end{align}


\subsection{Renormalization at leading order}

Loop diagrams will produce ultraviolet (UV) divergences that must be
regularized and removed through renormalization.
The divergences in the total EMT can be controlled and eliminated
through conventional renormalization of the QED Lagrangian.
We use dimensional regularization~\cite{Bollini:1972ui,tHooft:1972tcz,Collins:1984xc}
with $d = 4-2\epsilon$
to control these divergences.

The bare quantities appearing in Eqs.~(\ref{eqn:lagrangian:qed})
and (\ref{eqn:emt:qed}) are related to the renormalized
quantities~\cite{Itzykson:1980rh}:
\begin{subequations}
  \begin{align}
    \psi_0 &= \sqrt{Z_2} \psi
    \\
    m_0 &= Z_m m_e
    \\
    A^\mu_0 &= \sqrt{Z_3} A^\mu
    \\
    \lambda_0 &= Z_3^{-1} \lambda
    \\
    e_0 &= Z_e e
    \\
    Z_1 &\equiv \sqrt{Z_3} Z_2 Z_e
    \,.
  \end{align}
\end{subequations}
By the Ward identity, we have $Z_1=Z_2$ to all orders in
QED~\cite{Ward:1950xp,Takahashi:1957xn}.
Using these standard renormalized quantities,
the EMT takes the form:
\begin{multline}
  \label{eqn:EMT:photon:renormalized}
  T^{\mu\nu}(x)
  =
  Z_3 F^{\mu\lambda} F_{\lambda}^{\phantom{\lambda}\nu}
  +
  \frac{1}{4} g^{\mu\nu} Z_3 F^{\lambda\rho} F_{\lambda\rho}
  +
  Z_2 \bar{\psi}
  \left( \frac{i}{4}\overleftrightarrow{\partial}^{\{\mu}\gamma^{\nu\}} - Z_m m_e \right)
  \psi
  -
  \frac{1}{2} e Z_1 \bar{\psi} \gamma^{\{\mu}A^{\nu\}} \psi
  \\
  - \lambda (\partial\cdot A) \partial^{\{\mu} A^{\nu\}}
  + \frac{1}{2} g^{\mu\nu} \lambda (\partial\cdot A)^2
  \,.
\end{multline}
The values of the renormalization constants cannot be determined
without specifying a subtraction scheme.
For simplicity, we consider on-shell renormalization,
in which $Z_m$ is determined by requiring that $m_e$
be the physical pole mass of the electron,
and $Z_2$ and $Z_3$ are determined by requiring the residue of the renormalized
Green's functions are $1$ at the respective mass poles.
At leading order, the results for these constants are:
\begin{subequations}
  \begin{align}
    Z_m
    &=
    1
    +
    \frac{\alpha}{4\pi}
    \left\{
      -
      \frac{3}{\epsilon}
      +
      3 \log\left(\frac{m_e^2}{\LDR^2}\right)
      -
      5
      \right\}
    + \mathcal{O}(\alpha^2)
    \\
    Z_2
    &=
    1
    +
    \frac{\alpha}{4\pi}
    \left\{
      -
      \frac{1}{\epsilon}
      +
      3 \log\left(\frac{m_e^2}{\LDR^2}\right)
      -
      2  \log\left(\frac{\mu^2}{\LDR^2}\right)
      -
      5
      \right\}
    + \mathcal{O}(\alpha^2)
    \\
    Z_3
    &=
    1
    +
    \frac{\alpha}{3\pi}
    \left\{
      -
      \frac{1}{\epsilon}
      +
      \log\left(\frac{m_e^2}{\LDR^2}\right)
      \right\}
    + \mathcal{O}(\alpha^2)
  \end{align}
\end{subequations}
where $\LDR^2 = 4\pi \mu^2 e^{-\gamma_E}$.
Order $\alpha$ corrections to $Z_e$ are not necessary
for a leading order calculation of photon structure.

Lastly, before proceeding to calculations and results,
we specify that we use Feynman gauge ($\lambda=1$)
for ease of calculation.


\subsection{Leading-order calculations and results}

\begin{figure}
  \centering
  \includegraphics{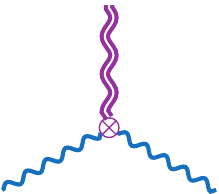}
  \includegraphics{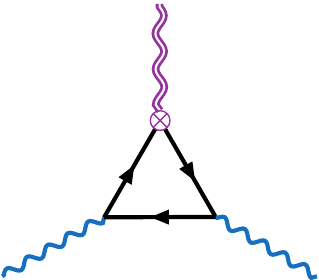}
  \includegraphics{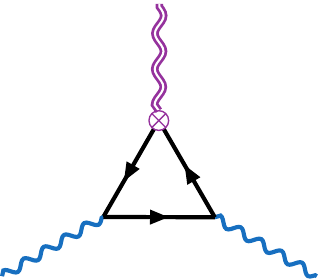}
  \includegraphics{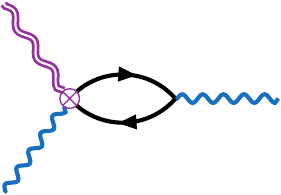}
  \includegraphics{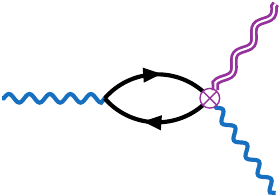}
  \caption{
    Feynman diagrams for leading-order QED corrections to
    the photon EMT.
    The EMT operator insertion is depicted as an
    interaction with a graviton.
  }
  \label{fig:diagrams}
\end{figure}

As discussed for instance in Ref.~\cite{Itzykson:1980rh},
the matrix element of the EMT vertex receives contributions from
truncated Feynman diagrams, i.e., those whose external legs
do not contain a self-energy part.
The relevant diagrams are depicted in Fig.~\ref{fig:diagrams},
where a graviton vertex is used to symbolize an EMT operator insertion,
as discussed in Ref.~\cite{Freese:2019bhb}.
\begin{align}
  \vcenter{ \hbox{ \includegraphics{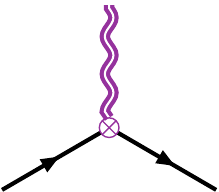} } }
  &=
  \frac{1}{2}
  Z_2
  \Big(
  k^\mu \gamma^\nu
  +
  k^\nu \gamma^\mu
  \Big)
  \\
  \vcenter{ \hbox{ \includegraphics{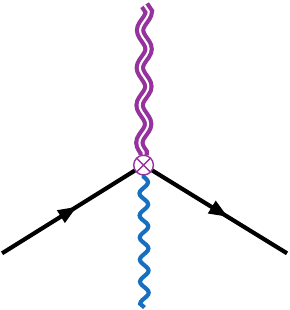} } }
  &=
  -\frac{e}{2}
  Z_1
  \Big(
  \gamma^\mu g^{\nu\lambda}
  +
  \gamma^\nu g^{\mu\lambda}
  \Big)
  \\
  \vcenter{ \hbox{ \includegraphics{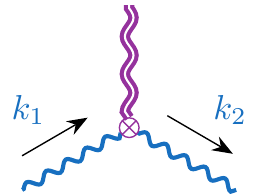} } }
  &=
  \begin{array}{l}
    - Z_3 \bigg\{
      \Big(k_1^\mu g^{\lambda\mu_1} - k_1^\lambda g^{\mu \mu_1}\Big)
      \Big(k_2^\nu g_{\lambda}^{\phantom{\lambda}\mu_2}
      - k_{2\lambda} g^{\nu \mu_2}\Big)
      + \Big( 1 \leftrightarrow 2 \Big)
      \bigg\}
    \\
    + \frac{1}{2} Z_3 g^{\mu\nu}
    \Big(k_1^\sigma g^{\lambda\mu_1} - k_1^\lambda g^{\sigma \mu_1}\Big)
    \Big(k_{2\sigma} g_{\lambda}^{\phantom{\lambda}\mu_2}
    - k_{2\lambda} g_{\sigma}^{\phantom{\sigma} \mu_2}\Big)
    \\
    - \lambda \bigg\{
      k_1^{\mu_1} \Big( k_2^\mu g^{\nu\mu_2} + k_2^\nu g^{\mu\mu_2} \Big)
      + \Big( 1 \leftrightarrow 2 \Big)
      \bigg\}
    + \lambda g^{\mu\nu} k_1^{\mu_1} k_2^{\mu_2}
    \,.
  \end{array}
\end{align}
Since the calculation is being done at leading order,
any $\mathcal{O}(\alpha^2)$ terms that arise will be dropped.

The first diagram in Fig.~\ref{fig:diagrams} (the direct diagram)
is straightforward:
\begin{align}
  T^{\mu\nu}_{\mathrm{dir}}
  \equiv
  \vcenter{ \hbox{ \includegraphics{direct.pdf} } }
  =
  Z_3
  \Theta_1^{\mu\nu}
  \,.
\end{align}
The last two diagrams (the eye diagrams) we find to be individually zero, so:
\begin{align}
  T^{\mu\nu}_{\mathrm{eye}}
  \equiv
  \vcenter{ \hbox{ \includegraphics{eye1.pdf} } }
  +
  \vcenter{ \hbox{ \includegraphics{eye2.pdf} } }
  =
  0
  \,.
\end{align}
The last two diagrams (the triangle diagrams)
are equal to each other,
and are constrained by consistency with
Eq.~(\ref{eqn:Milton}) to give, at leading order:
\begin{align}
  T^{\mu\nu}_{\mathrm{tri}}
  \equiv
  \vcenter{ \hbox{ \includegraphics{triangle1.pdf} } }
  +
  \vcenter{ \hbox{ \includegraphics{triangle2.pdf} } }
  =
  \big(F_1(t) - Z_3\big) \Theta_1^{\mu\nu}
  +
  F_2(t) \Theta_2^{\mu\nu}
  +
  F_3(t) \Theta_3^{\mu\nu}
  \,.
\end{align}
Our explicit results for $F_{1-3}(t)$ are given in
Eq.~(\ref{eqn:photon:gffs}) below.

In principle, it would have been possible for individual diagrams
to give additional tensor structures beyond those encountered
in Eq.~(\ref{eqn:Milton}), since that equation is constrained to satisfy
momentum conservation and individual diagrams are not.
However, it happens to work out that no non-conserving form factors
appear for individual diagrams,
in stark contrast to the case for an electron target~\cite{Rodini:2020pis}.
\added{
To be sure, the direct diagram cannot contain non-conserved form factors
since it only rescales the tree level result,
and since the eye diagrams are zero,
the triangle diagrams cannot contain
any new (non-conserved) tensor structures.
}

Summing the diagrams gives the following leading order results
for the gravitational form factors of the photon:
\begin{subequations}
  \label{eqn:photon:gffs}
  \begin{align}
    F_1(t)
    &=
    1
    +
    \frac{\alpha}{2\pi} \left\{
    \frac{35}{18}
    - \frac{13}{6} \frac{1}{\tau}
    -
    \frac{4}{3}
    \sqrt{\frac{1+\tau}{\tau}} \sinh^{-1}(\sqrt{\tau})
    \left( 1 - \frac{5}{4} \frac{1}{\tau} \right)
    -
    \frac{\big(\sinh^{-1}(\sqrt{\tau})\big)^2}{\tau}
    \left( 1 - \frac{1}{2}\frac{1}{\tau} \right)
    \right\}
    \\
    2t F_3(t)
    &=
    \frac{\alpha}{2\pi}
    \left\{
      \frac{1}{3}
      -
      7 \frac{1}{\tau}
      +
      6
      \frac{1}{\tau}
      \sqrt{\frac{1+\tau}{\tau}} \sinh^{-1}(\sqrt{\tau})
      -
      2 \frac{\big(\sinh^{-1}(\sqrt{\tau})\big)^2}{\tau}
      \left( 1 - \frac{1}{2} \frac{1}{\tau} \right)
      \right\}
    \\
    2t F_2(t)
    &=
    \frac{\alpha}{2\pi}
    \left\{
      \frac{1}{3}
      +
      3 \frac{1}{\tau}
      -
      2
      \frac{1}{\tau}
      \sqrt{\frac{1+\tau}{\tau}} \sinh^{-1}(\sqrt{\tau})
      -
      \frac{1}{\tau}
      \frac{\big(\sinh^{-1}(\sqrt{\tau})\big)^2}{\tau}
      \right\}
    \,,
  \end{align}
\end{subequations}
where $\tau = \frac{-t}{4m_e^2}$.
These results agree with those of Refs.~\cite{Berends:1975ah,Milton:1977je}.
In contrast to the individual diagrams, the form factors as a whole
are independent of renormalization scheme and scale---as expected,
since the EMT is a conserved current.

\begin{figure}
  \includegraphics[width=0.49\textwidth]{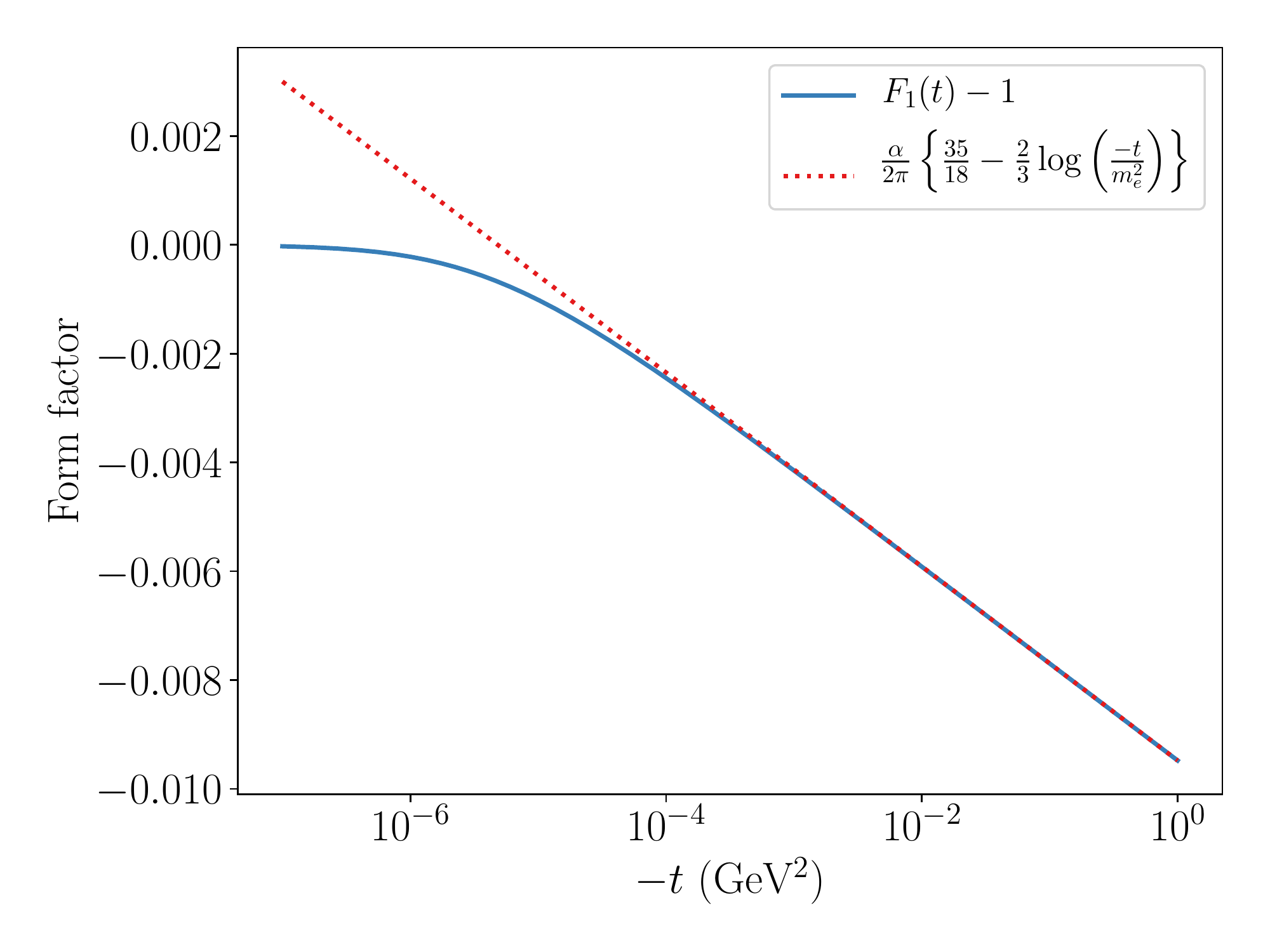}
  \includegraphics[width=0.49\textwidth]{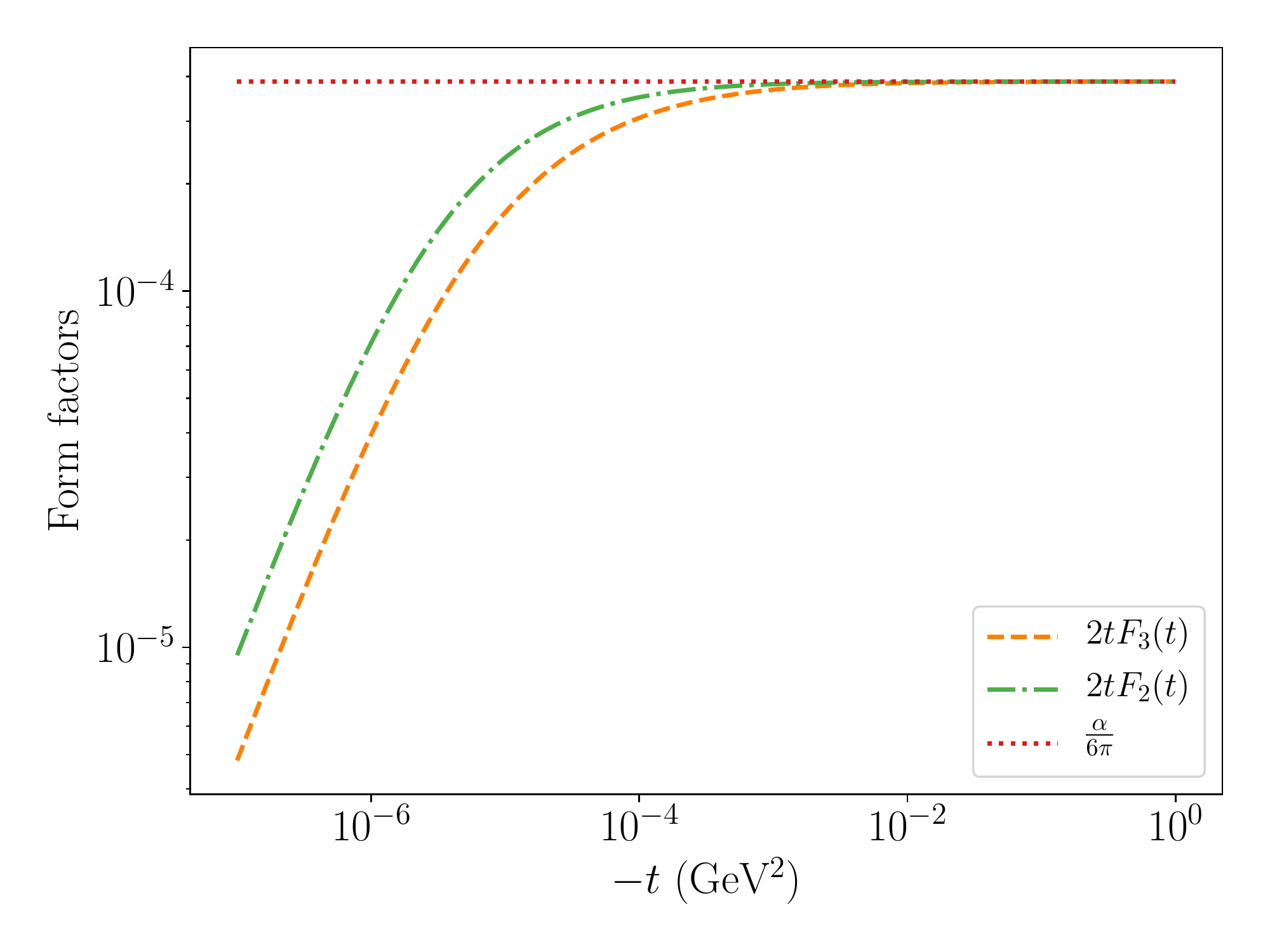}
  \caption{
    The gravitational form factors of the photon at leading order in QED,
    compared to their large $-t$ asymptotic forms.
  }
  \label{fig:photon:gffs}
\end{figure}

Numerical results for the gravitational form factors of the photon
are given in Fig.~\ref{fig:photon:gffs},
where they are also compared to the leading large $-t$ behavior given in
Eqs.~(\ref{eqn:photon:gffs:big}).

The limiting behavior of the form factors is instructive to consider,
since their behavior at small $-t$ is related to the size of the system
(for instance, via radii); and conversely, the behavior at large $-t$
is related to the behavior of the Fourier transform at small impact
parameters.
The small $-t$ expansions for each of these form factors are:
\begin{subequations}
  \label{eqn:photon:gffs:smol}
  \begin{align}
    \label{eqn:photon:gffs:smol:F1}
    F_1(t)
    &\approx
    1
    -
    \frac{\alpha}{2\pi}
    \frac{11}{45} \tau
    +
    \mathcal{O}(\tau^2)
    \\
    \label{eqn:photon:gffs:smol:F3}
    2t F_3(t)
    &\approx
    \frac{\alpha}{2\pi}
    \frac{2}{45} \tau
    +
    \mathcal{O}(\tau^2)
    \\
    2t F_2(t)
    &\approx
    \frac{\alpha}{2\pi}
    \frac{4}{45} \tau
    +
    \mathcal{O}(\tau^2)
    \,,
  \end{align}
\end{subequations}
  while the large $-t$ expansions are:
\begin{subequations}
  \label{eqn:photon:gffs:big}
  \begin{align}
    \label{eqn:photon:gffs:big:F1}
    F_1(t)
    &\approx
    1
    +
    \frac{\alpha}{2\pi}
    \left\{
      \frac{35}{18}
      -
      \frac{2}{3}
      \log(4\tau)
      +
      \frac{-10 + 2\log(4\tau) - \log^2(4\tau) }{4\tau}
      +
      \mathcal{O}\left(\frac{1}{\tau^2}\right)
      \right\}
    \\
    2t F_3(t)
    &\approx
    \frac{\alpha}{2\pi}
    \left\{
      \frac{1}{3}
      +
      \frac{-28 + 12\log(4\tau) - 2\log^2(4\tau) }{4\tau}
      +
      \mathcal{O}\left(\frac{1}{\tau^2}\right)
      \right\}
    \\
    2t F_2(t)
    &\approx
    \frac{\alpha}{2\pi}
    \left\{
      \frac{1}{3}
      +
      \frac{12 - 4\log(4\tau) }{4\tau}
      +
      \mathcal{O}\left(\frac{1}{\tau^2}\right)
      \right\}
    \,.
  \end{align}
\end{subequations}
The leading terms in the large $-t$ expansion agree with the results
previously obtained by Ref.~\cite{Milton:1977je}.

Noting the small $-t$ behavior in Eqs.~(\ref{eqn:photon:gffs:smol}),
the leading order corrections to the form factors
are clearly non-zero only at non-zero $t$,
so do not affect any static quantities.
As anticipated above, we find the following effective form factors
for photon states of any polarization:
\begin{subequations}
  \begin{align}
    \mathcal{A}(0) &= 1 \\
    \mathcal{J}(0) &= 1 \\
    \mathcal{D}(0) &= 1
    \,,
  \end{align}
\end{subequations}
just as in the free case.
For $\mathcal{A}(0)$ and $\mathcal{J}(0)$,
these results are constrained by conservation laws
(momentum and angular momentum, respectively).
Finding $\mathcal{D}(0)$ to be unaltered by leading order
corrections appears a non-trivial result,
but as discussed above, this follows from gauge invariance,
\replaced{
and introducing more interactions and higher orders will not
make $\mathcal{D}(0)$ negative for the photon.
}{
and introducing more interactions and higher orders will not save the
often postulated $\mathcal{D}(0) < 0$ stability condition.
}


\subsection{Electron-photon decomposition}

Let us now consider the breakdown into photon and electron contributions.
Of the diagrams in Fig.~\ref{fig:diagrams}, only the direct diagram
corresponds to ``photon'' contributions to the EMT.
Thus, without additional renormalization, we would have:
\begin{subequations}
  \begin{align}
    F_1^{(\gamma)}(t) &= Z_3
    \\
    F_1^{(e)}(t) &= -Z_3 + F_1(t)
    \,,
  \end{align}
\end{subequations}
but since $Z_3$ is UV divergent, both of these quantities are infinite.
Additional renormalization is required
for the composite operators defining the electron and photon contributions.

To simplify the renormalization procedure,
we follow Refs.~\cite{Ji:1995sv,Ji:2021qgo} in separating the EMT
into scalar $(0,0)$ and traceless $(1,1)$ parts,
with careful attention to this separation being done in
$d=4-2\epsilon$ dimensions:
\begin{subequations}
  \begin{align}
    T^{\mu\nu}
    &=
    \overline{T}^{\mu\nu}
    +
    \widehat{T}^{\mu\nu}
    \\
    \widehat{T}^{\mu\nu}
    &=
    \frac{g^{\mu\nu}}{4-2\epsilon}
    \,
    g_{\alpha\beta}\widehat{T}^{\alpha\beta}
    \\
    \overline{T}^{\mu\nu}
    &=
    T^{\mu\nu}
    -
    \widehat{T}^{\mu\nu}
    \,.
  \end{align}
\end{subequations}
As stated in Refs.~\cite{Ji:1995sv,Ji:2021qgo},
the scalar and traceless parts do not mix under renormalization,
so we can deal with them separately.
Moreover, since there are no non-conserved form factors,
and all of the structures $\Theta_{1,2,3}^{\mu\nu}$
defined in Eq.~(\ref{eqn:Milton}) contain $(1,1)$ pieces,
it is only necessary to consider
renormalization of the traceless piece of the EMT
to obtain form factor decompositions.

Therefore, let us consider how the traceless part of the EMT can be decomposed
into electron and photon pieces.
For the \emph{bare} operators:
\begin{subequations}
  \begin{align}
    \overline{T}^{\mu\nu}_{e0}
    &=
    \frac{i}{4} \bar{\psi}_0 \gamma^{\{\mu} \overleftrightarrow{\partial}^{\nu\}} \psi_0
    -
    \frac{1}{2} e_0 \bar{\psi}_0 \gamma^{\{\mu}_{\phantom{0}} A_0^{\nu\}} \psi_0
    -
    \frac{g^{\mu\nu}}{4-2\epsilon} m_{e0} \bar{\psi}_0 \psi_0
    \\
    \overline{T}^{\mu\nu}_{\gamma0}
    &=
    F_0^{\mu\lambda} F_{0\lambda}^{\phantom{0\lambda}\nu}
    +
    \frac{g^{\mu\nu}}{4-2\epsilon} F_0^2
    -
    \lambda_0 (\partial\cdot A_0)
    \left\{
      \partial^\mu A_0^\nu + \partial^\nu A_0^\mu
      -
      \frac{1-\epsilon}{2-\epsilon} g^{\mu\nu}
      (\partial\cdot A_0)
      \right\}
    \,.
  \end{align}
\end{subequations}
The renormalized operators
(which we signify by square brackets, following Ref.~\cite{Collins:1984xc})
are in general mixtures of the bare operators:
\begin{subequations}
  \begin{align}
    [\overline{T}^{\mu\nu}_e]
    &=
    Z_{ee}
    \overline{T}^{\mu\nu}_{e0}
    +
    Z_{e\gamma}
    \overline{T}^{\mu\nu}_{\gamma0}
    \\
    [\overline{T}^{\mu\nu}_\gamma]
    &=
    Z_{\gamma e}
    \overline{T}^{\mu\nu}_{e0}
    +
    Z_{\gamma\gamma}
    \overline{T}^{\mu\nu}_{\gamma0}
    \,,
  \end{align}
\end{subequations}
where the notation $Z_{ee}$ etc.\ comes from Ref.~\cite{Ji:1995sv},
and where gauge-fixing and equation of motion terms that vanish
for physical states have been dropped.

In the absence of interactions ($\alpha=0$),
we would have $Z_{ee} = Z_{\gamma\gamma} = 1$
and $Z_{e\gamma} = Z_{\gamma e} = 0$,
so it's helpful to define:
\begin{subequations}
  \begin{align}
    Z_{ee} &= 1 + \delta_{ee}
    \\
    Z_{\gamma\gamma} &= 1 + \delta_{\gamma\gamma}
    \,,
  \end{align}
\end{subequations}
where $\delta_{ee}$ and $\delta_{\gamma\gamma}$ are both
$\mathcal{O}(\alpha)$.
The additional requirement that
$[\overline{T}^{\mu\nu}] = \overline{T}^{\mu\nu}_0$~\cite{Ji:1995sv}
gives us:
\begin{subequations}
  \begin{align}
    Z_{\gamma e} &= - \delta_{ee}
    \\
    Z_{e\gamma} &= - \delta_{\gamma\gamma}
    \,.
  \end{align}
\end{subequations}
The divergent parts of $\delta_{ee}$ and $\delta_{\gamma\gamma}$
can be determined by requiring that matrix elements of the renormalized
operators are finite.
For the photon target at $\Delta=0$ in particular:
\begin{subequations}
  \begin{align}
    \langle \gamma |
    [\overline{T}^{\mu\nu}_{e}]
    | \gamma \rangle
    &=
    \left\{
      \frac{\alpha}{3\pi} \frac{1}{\epsilon}
      - \delta_{\gamma\gamma}
      \right\}
    \Theta_1^{\mu\nu}
    + \Big(\mathrm{finite}~\mathcal{O}(\alpha)\Big)
    \\
    \langle \gamma |
    [\overline{T}^{\mu\nu}_{\gamma}]
    | \gamma \rangle
    &=
    \left\{
      1 + \delta_{\gamma\gamma}
      -
      \frac{\alpha}{3\pi} \frac{1}{\epsilon}
      \right\}
    \Theta_1^{\mu\nu}
    + \Big(\mathrm{finite}~\mathcal{O}(\alpha)\Big)
    \,,
  \end{align}
\end{subequations}
which determines $\delta_{\gamma\gamma}$ at leading order in $\alpha$ to be:
\begin{align}
  \delta_{\gamma\gamma}
  &=
  \frac{\alpha}{3\pi}
  \frac{1}{\epsilon}
  +
  \alpha \mathscr{C}
  \,,
\end{align}
where $\mathscr{C}$ is a constant
that is defined by the renormalization scheme.
The other constant, $\delta_{ee}$, does not contribute to photon structure
at leading order, since it is $\mathcal{O}(\alpha)$ and appears
with another factor $\alpha$ in the relevant diagrams.

The finite electron-photon decomposition of the form factors
can now be performed by using the $\delta_{\gamma\gamma}$ result we have obtained.
Using Milton's basis for the form factors, only $F_1(t)$ receives additional
renormalization, and the decomposition can be written:
\begin{subequations}
  \begin{align}
    F_1^{(e)}(t;\mu^2)
    &=
    F_1(t) - 1
    +
    \frac{\alpha}{3\pi}
    \left\{
      \log\left(\frac{\mu^2}{m_e^2}\right)
      + \log(4\pi) - \gamma_E
      \right\}
    -
    \alpha \mathscr{C}
    \\
    F_1^{(\gamma)}(t;\mu^2)
    &=
    1
    -
    \frac{\alpha}{3\pi}
    \left\{
      \log\left(\frac{\mu^2}{m_e^2}\right)
      + \log(4\pi) - \gamma_E
      \right\}
    +
    \alpha \mathscr{C}
    \,.
  \end{align}
\end{subequations}
The decomposition is notably scale and scheme dependent.
By contrast, one has (at leading order):
\begin{subequations}
  \begin{align}
    F_2(t) &= F_2^{(e)}(t) \\
    F_3(t) &= F_3^{(e)}(t)
  \end{align}
\end{subequations}
which are finite as they are.

Let us consider a couple of schemes for illustration.
Note that on-shell subtraction was used for renormalization of the Lagrangian,
and that the following schemes are only used
for operator renormalization on top of this.
The following decompositions would differ if we had used a different
subtraction scheme when renormalizing the Lagrangian.

The first scheme is minimal subtraction (MS),
in which counterterms are defined only to cancel divergences,
and contain no finite part.
In this scheme, $\mathscr{C}_{\mathrm{MS}} = 0$.

The other scheme we consider is modified minimal subtraction
($\overline{\mathrm{MS}}$),
in which counterterms contain a finite part that cancels factors of
$\log(4\pi) - \gamma_E$ that show up frequently in dimensional regularization.
In this scheme, we have:
\begin{align}
  \mathscr{C}_{\overline{\mathrm{MS}}}
  =
  \frac{1}{3\pi}
  \Big\{
    \log(4\pi) - \gamma_E
    \Big\}
  \,.
\end{align}
An interesting aspect of the $\overline{\mathrm{MS}}$ scheme
for operator renormalization
(when combined with on-shell renormalization for the Lagrangian)
is that at a scale equal to the electron mass ($\mu = m_e$),
all of the photon target's light front momentum can be attributed
to the photon itself; i.e., one has:
\begin{subequations}
  \begin{align}
    \mathcal{A}_{\overline{\mathrm{MS}}}^{(e)}(0;m_e^2)
    &=
    0
    \\
    \mathcal{A}_{\overline{\mathrm{MS}}}^{(\gamma)}(0;m_e^2)
    &=
    1
    \,.
  \end{align}
\end{subequations}
This can only be true at a single renormalization scale, however,
since the momentum fractions obey an evolution
equation~\cite{Gribov:1972ri,Dokshitzer:1977sg,Altarelli:1977zs}.


\subsubsection{Trace piece of the EMT}

Although it is not necessary to obtain form factor decompositions,
it is interesting to look at the renormalization of the
pure trace part of the EMT.
The relevant bare operators are:
\begin{subequations}
  \begin{align}
    \widehat{T}^{\mu\nu}_{e0}
    &=
    \frac{g^{\mu\nu}}{4-2\epsilon}
    m_{e0} \bar\psi_0 \psi_0
    \\
    \widehat{T}^{\mu\nu}_{\gamma0}
    &=
    g^{\mu\nu}
    \left( \frac{1}{4} - \frac{1}{4-2\epsilon} \right)
    \Big(
    F_0^2
    -
    2 \lambda_0 (\partial\cdot A_0)^2
    \Big)
    \approx
    \frac{g^{\mu\nu}}{4}
    \frac{\epsilon}{2}
    \Big(
    F_0^2
    -
    2 \lambda_0 (\partial\cdot A_0)^2
    \Big)
    +
    \mathcal{O}(\epsilon^2)
    \,.
  \end{align}
\end{subequations}
The operator renormalization of the operators on the right hand side
are highly standardized
(see Refs.~\cite{Adler:1976zt,Collins:1976yq,Nielsen:1977sy,Tarrach:1981bi,Ji:1995sv,Hatta:2018sqd,Rodini:2020pis}).
The electron operator is a finite sigma term and is invariant under renormalization:
\begin{align}
  [m_e \bar\psi \psi]
  =
  m_{e0} \bar\psi_0 \psi_0
  \,,
\end{align}
and accordingly the $\epsilon$ dependence of its prefactor can be dropped.
The photon operator, however, mixes under renormalization:
\begin{align}
  \label{eqn:FF}
  [F^2]
  =
  Z_F F_0^2
  +
  Z_C
  m_{e0} \bar\psi_0 \psi_0
  \,.
\end{align}
The total renormalized EMT trace can be written:
\begin{align}
  [\widehat{T}^{\mu}_{\phantom{\mu}\mu}]
  =
  (1+\gamma_m) [m_e \bar\psi \psi]
  +
  \frac{\beta(e)}{2e} [F^2]
  \,.
\end{align}
As explained in Ref.~\cite{Rodini:2020pis,Metz:2020vxd},
it is somewhat arbitrary how these pieces can be attributed
to photon and electron contributions,
and a variety of schemes for breaking this up exist.

For the case of a photon target, however,
matrix elements of $[F^2]$ contribute at $\mathcal{O}(\alpha^2)$.
To start, $F_0^2$ itself evaluates to zero at leading order,
since only the direct diagram of Fig.~\ref{fig:diagrams}
could contribute.
Additionally, $Z_C$ is order $\alpha$,
and $m_{e0} \bar\psi_0 \psi_0$ only contributes through diagrams
that already have two electron-photon vertices.
Thus, each term on the right-hand side of Eq.~(\ref{eqn:FF})
is at least order $\alpha^2$ and does not contribute to
the leading order EMT matrix element.

Accordingly, for photon states at leading order:
\begin{align}
  \langle p' \lambda' | [\widehat{T}^{\mu\nu}] | p \lambda \rangle
  =
  \langle p' \lambda' | \widehat{T}^{\mu\nu}_{e0} | p \lambda \rangle
  +
  \mathcal{O}(\alpha^2)
  =
  \langle p' \lambda' | [\widehat{T}^{\mu\nu}_{e}] | p \lambda \rangle
  +
  \mathcal{O}(\alpha^2)
  \,.
\end{align}
Thus, the trace of the EMT for the photon at leading order
can be attributed entirely to the electron.
\added{
To be sure, the trace of the photon EMT vanishes at $t=0$,
both because $\Theta_1^{\mu\nu}$ is traceless
and because $\Theta_{2,3}^{\mu\nu}$ vanish at $t=0$.
Thus a mass sum rule based on the trace of the EMT
would be trivial for photons.
}

Since $\Theta_{2,3}^{\mu\nu}$ are not traceless,
the renormalization of the $(0,0)$ piece of the EMT
could conceivably affect the breakup of $F_{2,3}(t)$ into
electron and photon pieces.
However, we find that these form factors can be attributed
entirely to the electron at leading order,
consistent with our finding when considering
the renormalization of the $(1,1)$ piece.


\subsection{Radii and densities: helicity states}

As discussed in Refs.~\cite{Lorce:2018egm,Freese:2021czn}
as well as \companion,
two-dimensional densities of the photon in the transverse plane
can be obtained through 2D Fourier transforms of its
gravitational form factors.
\deleted{
The ability to describe densities of the photon in particular is
an opportunity provided specifically by the light front formalism;
since the photon does not have a rest frame,
the Breit frame formalism
(which purports to ascribe rest frame densities to particles)
is inapplicable.
}


\subsubsection{Light front momentum density}

Let us first consider the $P^+$ density.
For helicity states, this is given by
Eq.~(\ref{eqn:p+:hel})
\deleted{of \companion,}
with $\mathcal{A}(t) = F_1(t)$.
    In particular, the 2D Fourier transform of an
azimuthally symmetric function such as $F_1(t)$ can be written:
\begin{align}
  \rho_{P^+}(b_\perp)
  =
  \frac{P^+}{2\pi}
  \int_0^\infty \mathrm{d}k \,
  k F_1(-k^2)
  J_0(b_\perp k)
  =
  \frac{P^+}{2\pi}
  \mathscr{H}_0\Big[
    F_1(-k^2)
    \Big](b_\perp)
  \,,
\end{align}
where $\mathscr{H}_\nu[F(k)](b)$ signifies the Hankel transform of order $\nu$.
Analytic results for the Hankel transforms of the functions
in Eq.~(\ref{eqn:photon:gffs}) do not exist in the mathematics literature,
but a numerical Hankel transform can be used to obtain the densities,
provided the growing and constant asymptotic behavior at large $-t$
(as given in the leading terms of Eq.~(\ref{eqn:photon:gffs:big:F1}))
is subtracted off.
The two-dimensional Fourier transforms of the functions describing
this asymptotic behavior are as follows:
\begin{align}
  \int \frac{\mathrm{d}^2 \boldsymbol{\Delta}_\perp}{(2\pi)^2}
  e^{-i\mathbf{b}_\perp\cdot\boldsymbol{\Delta}_\perp}
  &=
  \delta^{(2)}(\mathbf{b}_\perp)
  \\
  \int \frac{\mathrm{d}^2 \boldsymbol{\Delta}_\perp}{(2\pi)^2}
  \log\left(\frac{\mathbf{\Delta}_\perp^2}{m_e^2}\right)
  e^{-i\mathbf{b}_\perp\cdot\boldsymbol{\Delta}_\perp}
  &=
  -\frac{ \Theta(m_e b_\perp) }{\pi b_\perp^2}
  \,,
\end{align}
where $\Theta(x)$ is the Heaviside step function.
The first of these identities is well-known;
the second follows from Eq.~(\ref{eqn:hankel:log}),
which is proved in the Appendix.

\begin{figure}
  \includegraphics[width=0.49\textwidth]{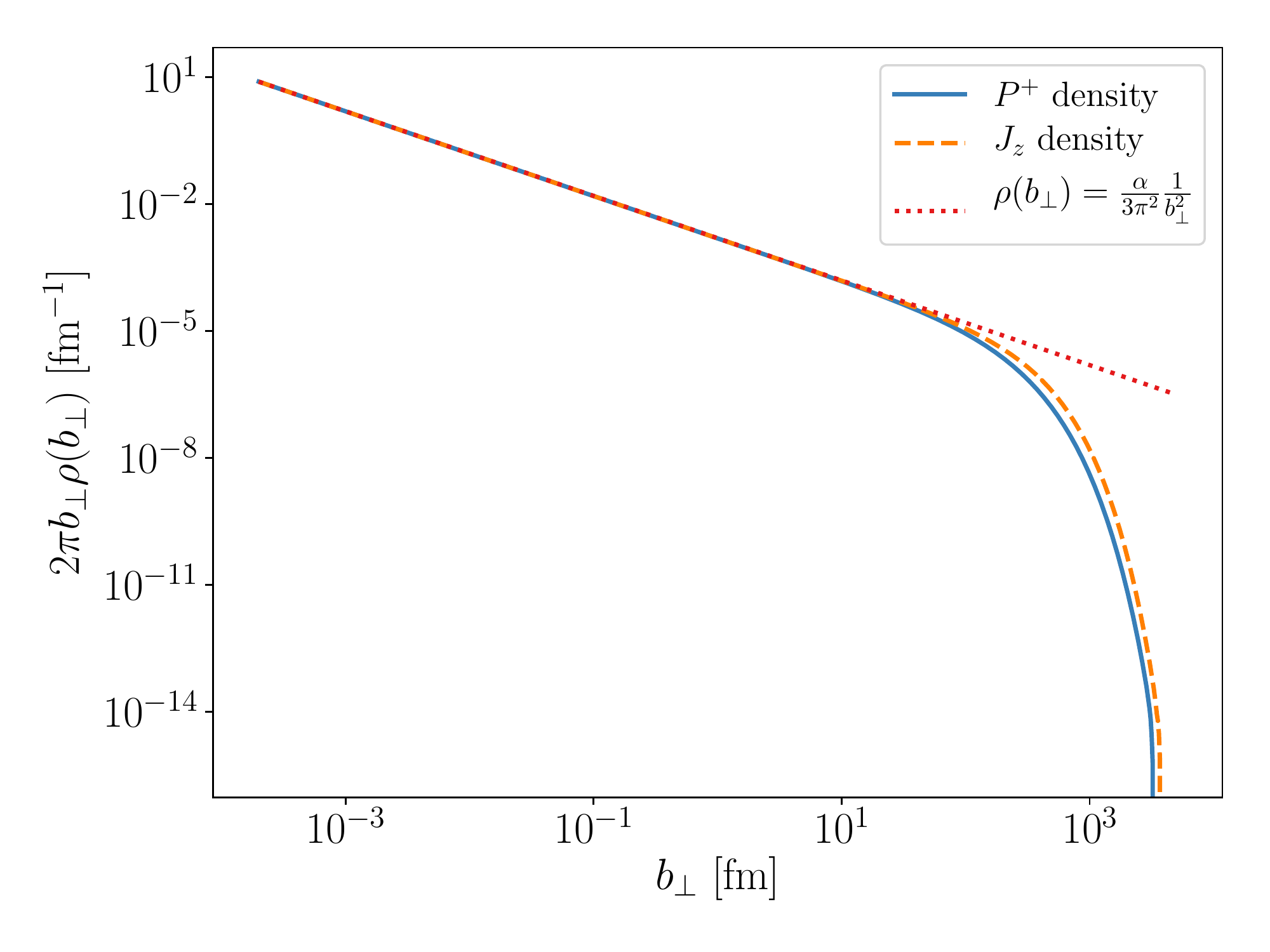}
  \includegraphics[width=0.49\textwidth]{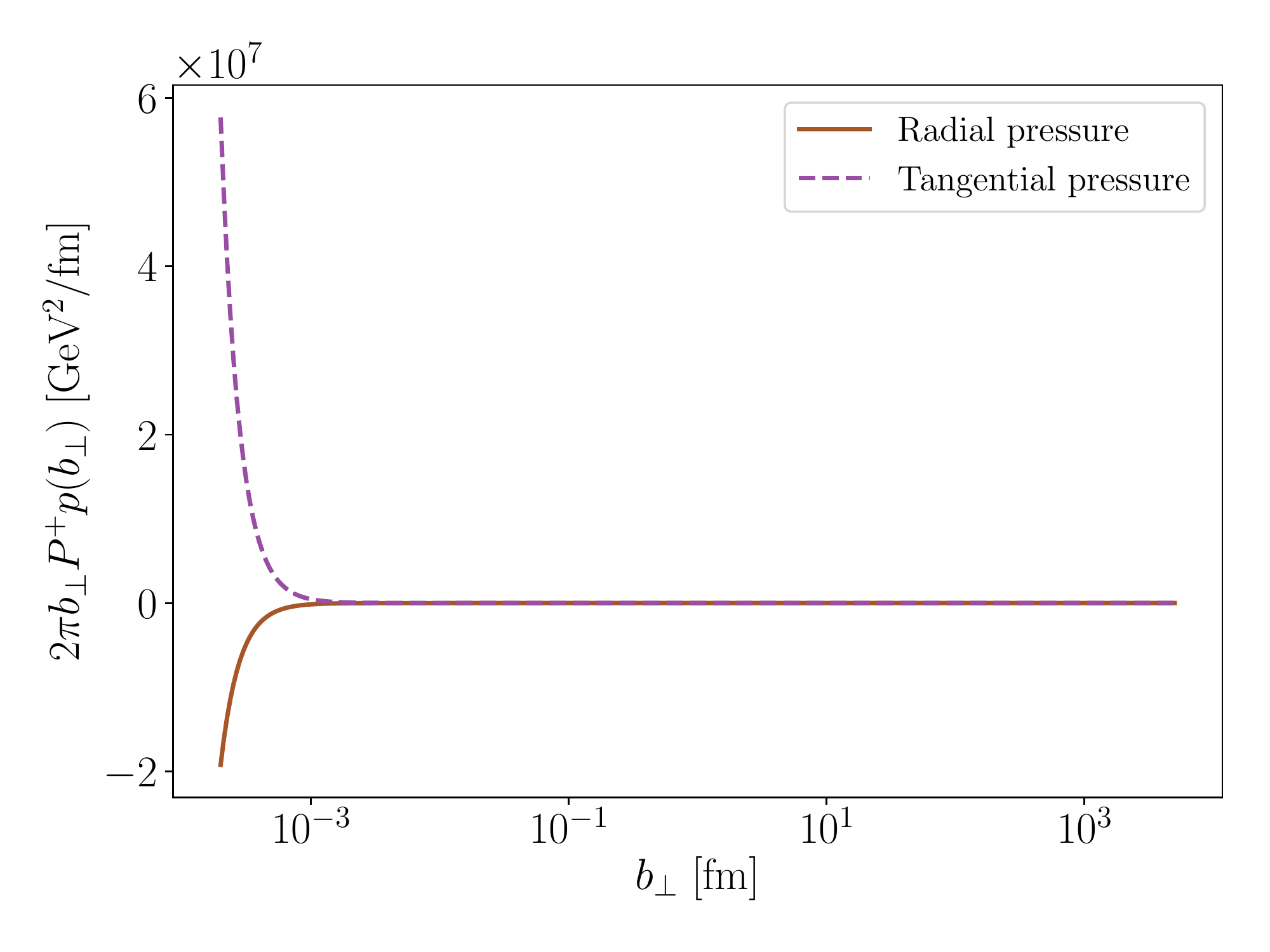}
  \caption{
    The light front densities of a helicity state photon.
    For the $P^+$ density, we have divided out $P^+$ to make
    the quantity boost-invariant,
    and so that it could be compared to the $J_z$ density.
    The pressure distributions have simiarly been multiplied by
    $P^+$ to produce boost-invariant quantities.
  }
  \label{fig:photon:densities}
\end{figure}

Since the large $-t$ behavior of the form factor governs the
small $b_\perp$ behavior of the density, we can expect
the $P^+$ density at small (but non-zero) impact parameter to take the form:
\begin{align}
  \label{eqn:photon:p+:smol}
  \rho_{P^+}(b_\perp)
  \approx
  \frac{\alpha P^+}{3\pi^2} \frac{1}{b_\perp^2}
  +
  \mathcal{O}(b_\perp^{-1})
  \,,
\end{align}
with the $b_\perp^{-2}$ behavior coming specifically from
the Fourier transform of the logarithm in Eq.~(\ref{eqn:photon:gffs:big:F1}).
In the left panel of Fig.~\ref{fig:photon:densities},
the exact $P^+$ density is compared to its limiting form at small impact
parameter as given in Eq.~(\ref{eqn:photon:p+:smol})---as well as to
the $J_z$ density, which shall be described in more depth below.

The $P^+$ radius is given by the derivative of the form factor,
$4F_1'(0)$, as discussed for instance in \companion.
The value can be read off from the small $-t$ expansion
in Eq.~(\ref{eqn:photon:gffs:smol:F1}):
\begin{align}
  \langle b_\perp^2 \rangle_{P^+}
  =
  \frac{11}{90\pi}
  \frac{\alpha}{m_e^2}
  \approx
  (6.5~\mathrm{fm})^2
  \,.
\end{align}
This especially large radius is due to the photon easily forming
a large-sized configuration as an $e^+e^-$ pair,
owing to the small mass of the electron.
This large radius can also be understood through the small $b_\perp$
form of Eq.~(\ref{eqn:photon:p+:smol})---which is a fairly slow
falloff---holding quite well up to about $b_\perp \approx 10$~fm,
as seen in the left panel of Fig.~\ref{fig:photon:densities}.


\subsubsection{Angular momentum density}

As explicated in Eq.~(\ref*{P1:eqn:J:hel}) of \companion,
\begin{align}
  \label{eqn:J:hel}
  \rho_{J_z}^{(\lambda)}(b_\perp)
  &=
  \lambda
  \int \frac{\mathrm{d}^2\boldsymbol{\Delta}_\perp}{(2\pi)^2}
  \left\{
    \mathcal{J}(t)
    +
    t \frac{\mathrm{d} \mathcal{J}(t)}{\mathrm{d}t}
    \right\}
  e^{-i\boldsymbol{\Delta}_\perp\cdot\mathbf{b}_\perp}
  \,.
\end{align}
\deleted{
the angular momentum density is given by the Fourier transform
of $\mathcal{J}(t)+t\mathcal{J}'(t)$.
}
For a $\lambda=+1$ helicity state, $\mathcal{J}(t)=F_1(t)$.
It is helpful to note that:
\begin{align}
  F_1(t)
  +
  t \frac{\mathrm{d}F_1(t)}{\mathrm{d}t}
  =
  1
  +
  \frac{\alpha}{2\pi} \left\{
    \frac{23}{18}
    +
    \frac{5}{6}
    \frac{1}{\tau}
    -
    \frac{1}{3}
    \sqrt{\frac{\tau}{1+\tau}}
    \sinh^{-1}(\sqrt{\tau})
    \left(
    4 + \frac{5}{\tau} + \frac{1}{\tau^2}
    \right)
    -
    \frac{1}{2}
    \frac{\big( \sinh^{-1}(\sqrt{\tau}) \big)^2}{\tau^2}
    \right\}
  \,,
\end{align}
where $\tau = \frac{-t}{4m_e^2}$ as before.
The small $-t$ expansion is given by:
\begin{align}
  F_1(t)
  +
  t \frac{\mathrm{d}F_1(t)}{\mathrm{d}t}
  \approx
  1
  -
  \frac{\alpha}{2\pi}
  \frac{22}{45}
  \tau
  +
  \mathcal{O}(\tau^2)
\end{align}
and the large $-t$ limiting form by:
\begin{align}
  \label{eqn:J:big}
  F_1(t)
  +
  t \frac{\mathrm{d}F_1(t)}{\mathrm{d}t}
  \approx
  \frac{\alpha}{2\pi}
  \left\{
    \frac{23}{18}
    -
    \frac{2}{3}
    \log(4\tau)
    +
    \frac{ 2 - 2\log(4\tau) }{4\tau}
    \right\}
  +
  \mathcal{O}(\tau^{-2})
  \,.
\end{align}
The small $-t$ expansion tells us that the angular momentum radius is:
\begin{align}
  \langle b_\perp^2 \rangle_{J}
  =
  \frac{11}{45\pi}
  \frac{\alpha}{m_e^2}
  \approx
  (13~\mathrm{fm})^2
  \,,
\end{align}
which is twice the $P^+$ radius.

The coefficient attached to the logarithm in Eqs.~(\ref{eqn:photon:gffs:big:F1})
and (\ref{eqn:J:big}) is the same,
and accordingly the $P^+$ and $J_z$ densities will have the same small $b_\perp$
asymptotics.
This can be seen clearly in the numerical result in the left panel
of Fig.~\ref{fig:photon:densities}.
The larger $J_z$ radius must be attributed to the difference between the
$P^+$ and $J_z$ densities at larger $b_\perp$,
where the angular momentum density becomes larger.


\subsubsection{Pressure distributions}

The Polyakov stress potential $\widetilde{\mathcal{D}}(b_\perp)$
is related to the Fourier transform of $\mathcal{D}(t)$,
and since $\mathcal{D}(t)=\mathcal{A}(t)=F_1(t)$ for a helicity state photon,
we effectively have:
\begin{align}
  \widetilde{\mathcal{D}}(b_\perp)
  =
  \frac{1}{4(P^+)^2}
  \rho_{P^+}(b_\perp)
  \,.
\end{align}
The radial and tangential eigenpressures can be obtained from this stress potential
through Eqs.~(\ref{eqn:prt:hel}).
The derivatives can be done numerically for the exact densities,
but analytically for the limiting forms at small $b_\perp$.
These limiting forms are:
\begin{subequations}
  \begin{align}
    p_r(b_\perp)
    &\approx
    -
    \frac{\alpha}{6\pi^2 P^+} \frac{1}{b_\perp^4}
    +
    \mathcal{O}(b_\perp^{-3})
    \\
    p_t(b_\perp)
    &\approx
    +
    \frac{\alpha}{2\pi^2 P^+} \frac{1}{b_\perp^4}
    +
    \mathcal{O}(b_\perp^{-3})
    \,.
  \end{align}
\end{subequations}

Numerical results for the eigenpressures in a helicity state photon
are plotted in the right panel of Fig.~\ref{fig:photon:densities}.
Because of the $b_\perp^{-4}$ behavior,
these pressures become highly singular near the transverse origin.
The magnitude of the actual pressures depends on $P^+$
(being inversely proportional to it),
but a boost-invariant quantity can be obtained by multiplying
the pressures by $P^+$ (as is done in the plot).

The radial pressure, in contrast to the usual expectations
(of e.g.\ Refs.~\cite{Perevalova:2016dln,Polyakov:2018zvc,Lorce:2018egm,Freese:2021czn}),
is not strictly positive, but in fact highly negative.
This is, however, in line with our findings for the free photon.
\deleted{
and more strongly forces the conclusion that positive radial pressure
cannot be a universal stability criterion.
}
The negative pressure for the QED photon now has spatial extent,
rather than being localized at the origin through a delta function.
Despite the spatial extent, however, the mechanical radius remains zero.
Considering Eq.~(\ref*{P1:eqn:r:mech}) of \companion:
\begin{align}
  \langle b_\perp^2 \rangle_{\mathrm{mech}}
  =
  \frac{
    \int \mathrm{d^2}\mathbf{b}_\perp \,
    \mathbf{b}_\perp^2
    p_r(\mathbf{b_\perp})
  }{
    \int \mathrm{d^2}\mathbf{b}_\perp \,
    p_r(\mathbf{b_\perp})
  }
  \,,
\end{align}
$F_1(0)$---which appears in the numerator---is still finite,
but the denominator (which integrates $F_1(t)$ over all $t$)
is infinite.


\subsection{Linear polarization and density modulations}

Photons in superpositions of helicity states have
azimuthal modulations in their densities.
The magnitudes of these modulations are governed by
$F_2(t)$ (for stress distributions) and $F_3(t)$ (for $P^+$ densities).
We consider linear polarization here as an extreme case.
Taking the appropriate 2D Fourier transforms
of Eq.~(\ref{eqn:gffs:linear}) gives:
\begin{subequations}
  \begin{align}
    \rho_{P^+}^{\mathrm{linear}}(\mathbf{b}_\perp)
    &=
    \rho_{P^+}^{\mathrm{circular}}(b_\perp)
    +
    \cos 2\phi \,
    \frac{P^+}{2\pi}
    \mathscr{H}_2 \left[
      -2 k^2 F_3(-k^2)
      \right](b_\perp)
    \\
    \widetilde{D}^{\mathrm{linear}}(\mathbf{b}_\perp)
    &=
    \widetilde{D}^{\mathrm{circular}}(b_\perp)
    +
    \cos 2\phi \,
    \frac{1}{2\pi}
    \frac{1}{4P^+}
    \mathscr{H}_2 \left[
      -2 k^2 F_2(-k^2)
      \right](b_\perp)
  \end{align}
\end{subequations}
for the $P^+$ density and Polyakov stress potential.
The stress tensor can be decomposed into three functions,
as described by Eq.~(\ref*{P1:eqn:Sij:trans}) of \companion:
\begin{align}
  \label{eqn:Sij:trans}
  S^{ij}_T(\mathbf{b}_\perp, m_s)
  =
  \delta^{ij} p_T^{(m_s)}(\mathbf{b}_\perp)
  +
  \left( \hat{b}^i \hat{b}^j - \frac{1}{2} \delta^{ij} \right)
  s_T^{(m_s)}(\mathbf{b}_\perp)
  +
  \Big( \hat{b}^i \hat{\phi}^j + \hat{\phi}^i \hat{b}^j \Big)
  v_T^{(m_s)}(\mathbf{b}_\perp)
  \,.
\end{align}
The relevant functions can be shown
(with a little calculus and Bessel function identities) to be:
\begin{subequations}
  \label{eqn:photon:psv}
  \begin{align}
    p_{\mathrm{linear}}(\mathbf{b}_\perp)
    &=
    p_{\mathrm{circular}}(b_\perp)
    +
    \cos 2\phi \,
    \frac{1}{2\pi}
    \frac{1}{8P^+}
    \mathscr{H}_2 \left[
      -2 k^4 F_2(-k^2)
      \right](b_\perp)
    \\
    s_{\mathrm{linear}}(\mathbf{b}_\perp)
    &=
    s_{\mathrm{circular}}(b_\perp)
    +
    \cos 2\phi \,
    \frac{1}{2\pi}
    \frac{1}{8P^+}
    \Big\{
      \mathscr{H}_0 \left[
        -2 k^4 F_2(-k^2)
        \right](b_\perp)
      +
      \mathscr{H}_4 \left[
        -2 k^4 F_2(-k^2)
        \right](b_\perp)
      \Big\}
    \\
    v_{\mathrm{linear}}(\mathbf{b}_\perp)
    &=
    \sin 2\phi \,
    \frac{1}{2\pi}
    \frac{1}{8P^+}
    \Big\{
      -
      \mathscr{H}_0 \left[
        -2 k^4 F_2(-k^2)
        \right](b_\perp)
      +
      \mathscr{H}_4 \left[
        -2 k^4 F_2(-k^2)
        \right](b_\perp)
      \Big\}
    \,.
  \end{align}
\end{subequations}
From these, the eigenpressures of the linearly polarized photon
can be obtained using Eq.~(\ref*{P1:eqn:eigenpressure}) of \companion:
\begin{subequations}
  \label{eqn:eigenpressure}
  \begin{align}
    P_{T,\pm}^{(m_s)}(\mathbf{b}_\perp)
    &=
    p_T^{(m_s)}(\mathbf{b}_\perp)
    \pm
    \sqrt{
      \frac{1}{4} \big(s_T^{(m_s)}(\mathbf{b}_\perp)\big)^2
      +
      \big(v_T^{(m_s)}(\mathbf{b}_\perp)\big)^2
    }
    \\
    \theta_\pm^{(m_s)}(\mathbf{b}_\perp)
    &=
    \phi
    +
    \frac{1}{2}
    \tan^{-1}\left(
    \frac{ 2 v_T^{(m_s)}(\mathbf{b}_\perp) }{ s_T^{(m_s)}(\mathbf{b}_\perp) }
    \right)
    +
    \Theta\Big(\pm s_T^{(m_s)}(\mathbf{b}_\perp)\Big)
    \frac{\pi}{2}
    \\
    \hat{e}_\pm^{(m_s)}(\mathbf{b}_\perp)
    &=
    \cos\big(\theta_\pm^{(m_s)}(\mathbf{b}_\perp)\big)
    \, \hat{x}
    +
    \sin\big(\theta_\pm^{(m_s)}(\mathbf{b}_\perp)\big)
    \, \hat{y}
    \,.
  \end{align}
\end{subequations}

\begin{figure}
  \includegraphics[width=0.49\textwidth]{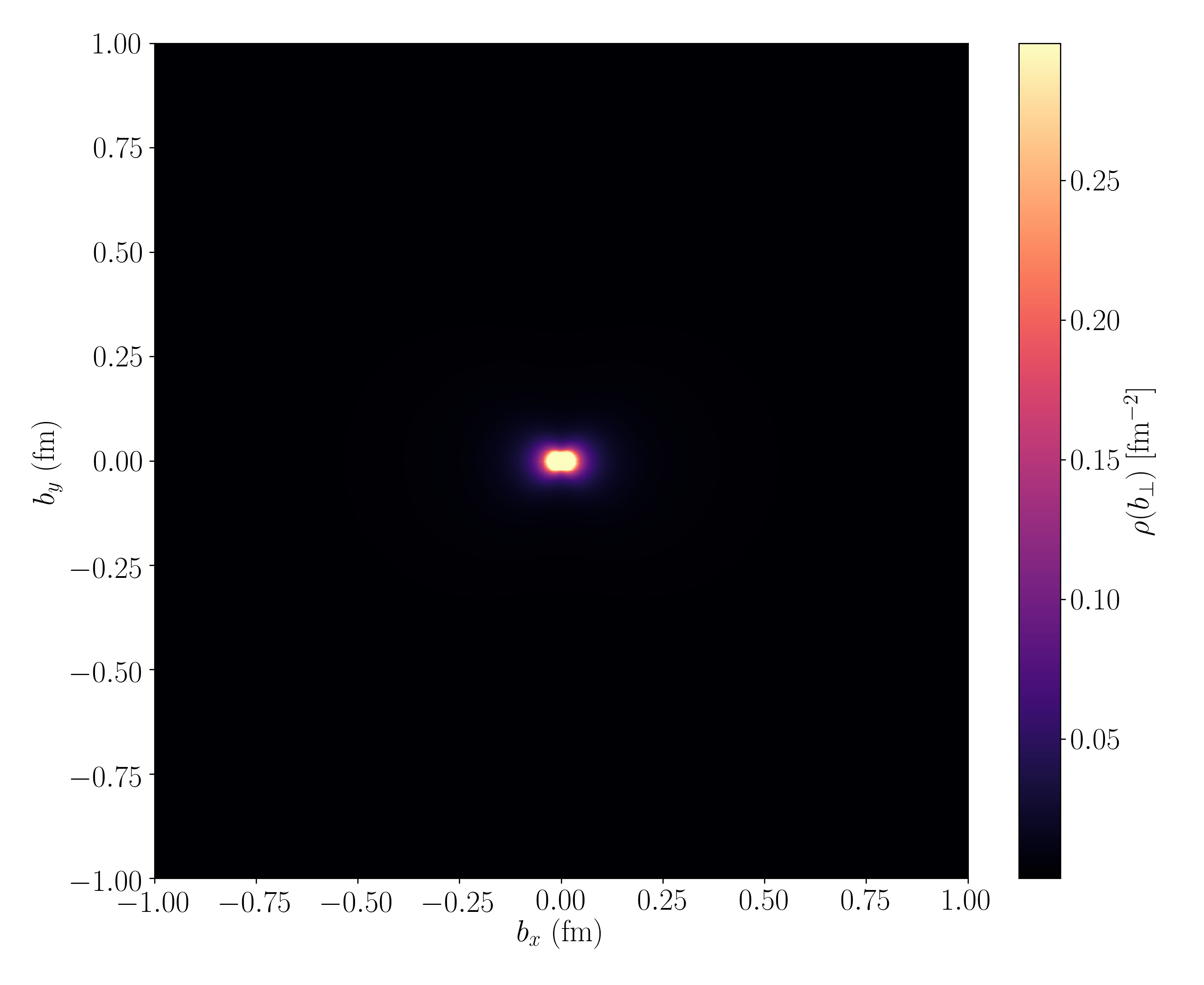}
  \includegraphics[width=0.49\textwidth]{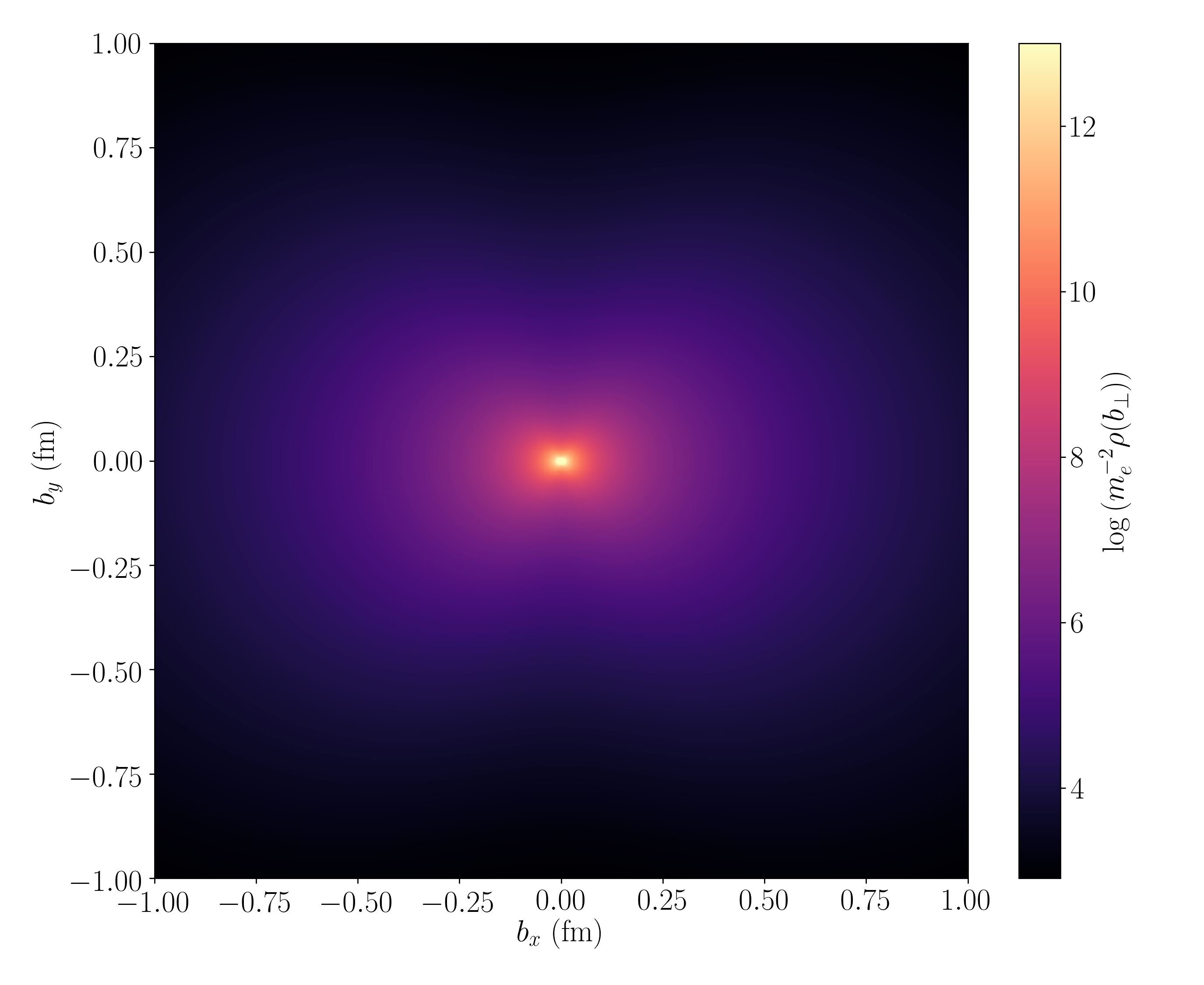}
  \caption{
    $P^+$ density of a horizontally polarized photon.
    The left panel uses a linear color map,
    but clips values above the maximum depicted in the color bar.
    The right panel uses a logarithmic color map
    to make the gradient less steep,
    and values are still clipped above the maximum
    depicted in the color bar.
    Both densities have had $P^+$ divided out to make them
    boost invariant.
  }
  \label{fig:photon:2D}
\end{figure}

Numerical results for the $P^+$ density of a horizontally polarized photon
are depicted in Fig.~\ref{fig:photon:2D}.
Because of the $\sim b_\perp^{-2}$ behavior of the density at small impact parameter,
the gradient is exceedingly steep in these plots,
and the density values above an arbitrary maximum have been clipped
to prevent the plots from displaying as an isolated white pixel.
The left panel, which uses a linear color map, clearly shows that the photon
is effectively elongated in the direction of polarization
(i.e., the direction of the electric field oscillations).
The right panel uses a logarithmic color map to make
the gradient more visible.

Since the $P^+$ density is not azimuthally symmetric,
there is a quadrupole moment associated with it.
This moment can be evaluated using Eq.~(\ref*{P1:eqn:QLF}) of \companion,
but with the linear polarization direction (in this case, $\hat{x}$)
used instead of a spin vector.
In terms of the form factors,
the quadrupole moment works out to be:
\begin{align}
  \mathcal{Q}_{\mathrm{LF}}
  =
  -8 F_3(0)
  \,.
\end{align}
As can be seen in Eq.~(\ref{eqn:photon:gffs:smol:F3}),
$F_3(0)$ is finite, and in particular:
\begin{align}
  \mathcal{Q}_{\mathrm{LF}}
  =
  \frac{\alpha}{45\pi}
  \frac{1}{m_e^2}
  \approx
  7.7~\mathrm{fm}^2
  \,.
\end{align}
This is a remarkably large quadrupole moment,
and is positive, indicating that the photon is prolate in the direction
of polarization.
This is of course compatible with what is visible in Fig.~\ref{fig:photon:2D}.

\begin{figure}
  \includegraphics[width=0.49\textwidth]{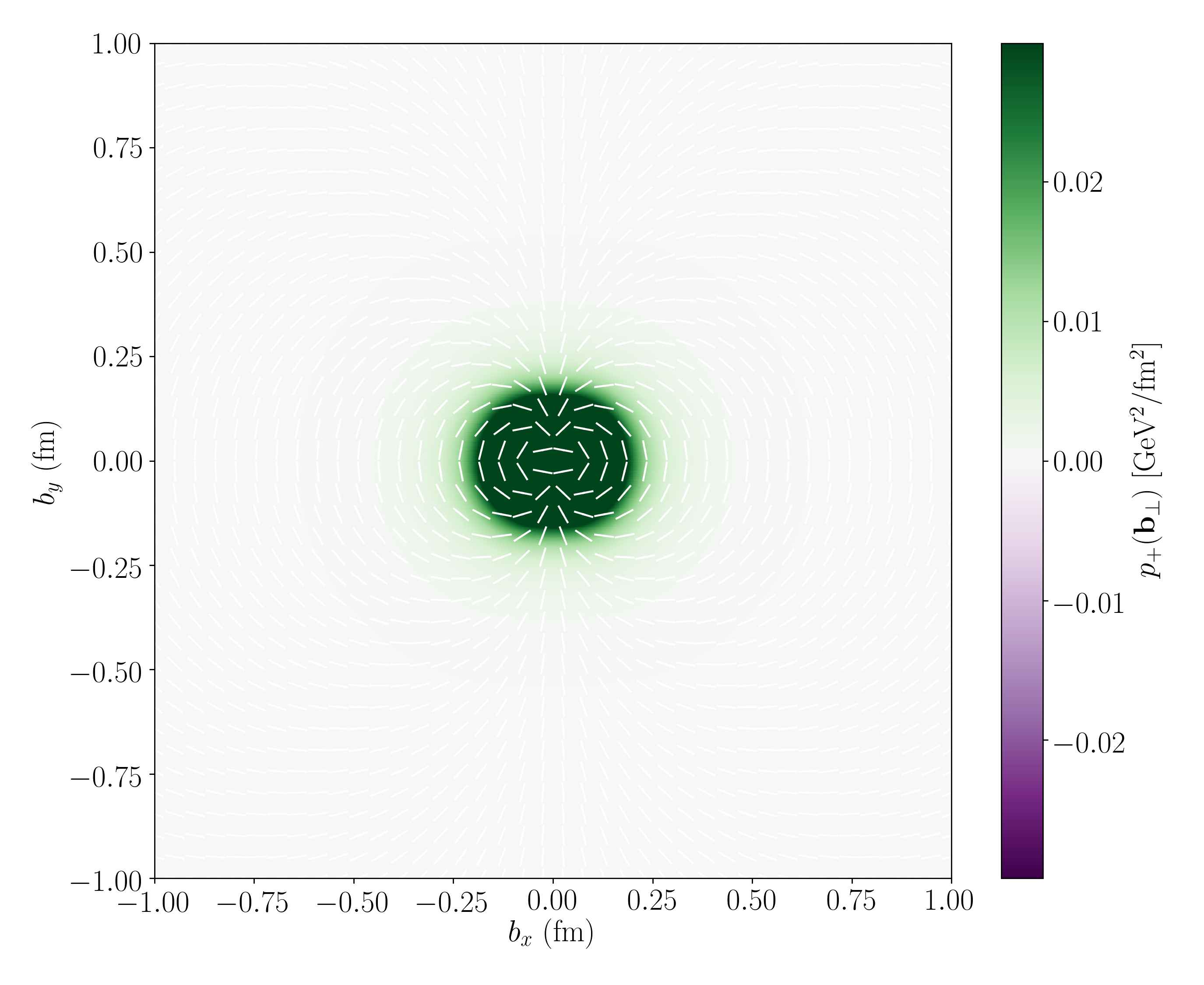}
  \includegraphics[width=0.49\textwidth]{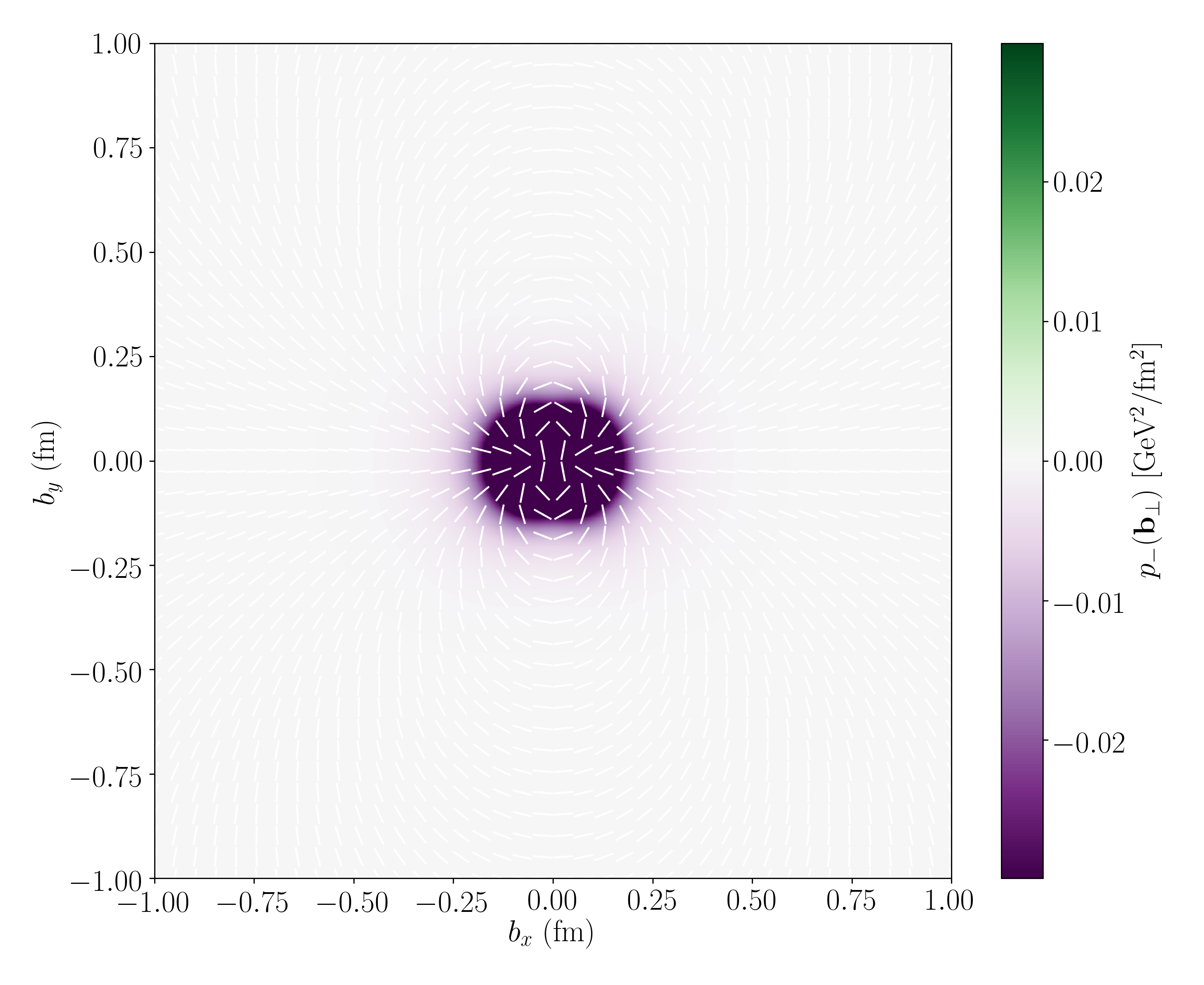}
  \caption{
    Eigenpressures of a horizontally polarized photon.
    The color maps have been clipped above the shown maximum and minimum values.
  }
  \label{fig:photon:2D:stress}
\end{figure}

Using Eq.~(\ref{eqn:photon:psv}),
along with Eq.~(\ref{eqn:eigenpressure}),
\deleted{of \companion,}
the eigenpressures of a horizontally polarized photon can be obtained.
Numerical results for these eigenpressures
are depicted in Fig.~\ref{fig:photon:2D:stress}.
In contrast to transversely polarized states of massive hadrons,
neither of these eigenpressures can be interpreted as a deformed radial
or tangential pressure.


\section{Radiation pressure}
\label{sec:rad}

For both the free photon and the QED photon at leading order,
we found a negative radial light front pressure.
This is a strange result that appears to contradict
our intuition (and known results~\cite{Jackson:1998nia})
that photons should exert positive pressure.
However, this contradiction is only apparent.
Radiation pressure---the pressure exerted \emph{by} photons---includes
contributions from the total \added{transverse} motion of the electromagnetic field,
which is to say it contains contributions from $\mathcal{A}(t)$
that are neglected by looking at only the \added{transversely} comoving part of the stress tensor.
(See Ref.~\cite{Freese:2021czn}, especially Sec.~III thereof, for further explanation.)

Let us consider the stress tensor as a whole.
As explained in Ref.~\cite{Freese:2021czn},
this diverges for states that are localized at the transverse origin.
Wave packets with a finite spatial extent must be used to define such a density,
which cannot be interpreted as describing intrinsic structure
(in contrast to the comoving stress tensor).
Let us consider a Gaussian wave packet with average \added{transverse} momentum
$\mathbf{k}_\perp$ and a finite transverse spatial width $\sigma$:
  \begin{align}
    \langle p^+, \mathbf{p}_\perp, \lambda | \Psi \rangle
    =
    \sqrt{2\pi} (2\sigma)
    e^{-\sigma^2(\mathbf{p}_\perp-\mathbf{k}_\perp)^2}
    \sqrt{2p^+(2\pi) \delta(p^+ - P^+)}
    \,.
  \end{align}
  Using Eq.~(6) of Ref.~\cite{Freese:2021czn} with
  the local operator $\hat{\mathcal{O}} = T^{ij}$,
  and noting the breakdown of the EMT in terms of helicity amplitudes in
  Eq.~(\ref{eqn:emt:spin1}), we find:
  \begin{align}
    T^{ij}(\mathbf{b}_\perp;\mathbf{k}_\perp,\sigma)
    &=
    \frac{ 1 }{2P^+}
    \int \frac{\mathrm{d}^2\mathbf{P}_\perp}{(2\pi)^2}
    \int \frac{\mathrm{d}^2\boldsymbol{\Delta}_\perp}{(2\pi)^2}
    \Bigg\{
      2
      \mathbf{P}_\perp^i
      \mathbf{P}_\perp^j
      \mathcal{A}(t)
      +
      \frac{1}{2}
      \Big(
      \boldsymbol{\Delta}_\perp^i \boldsymbol{\Delta}_\perp^j
      -
      \delta^{ij} \boldsymbol{\Delta}_\perp^2
      \Big)
      \mathcal{D}(t)
      \Bigg\}
    e^{-i\boldsymbol{\Delta}_\perp\cdot\mathbf{b}_\perp}
    e^{-2\frac{\sigma^2}{2} (\mathbf{P}-\mathbf{k})_\perp^2}
    e^{-\frac{\sigma^2}{2} \boldsymbol{\Delta}_\perp^2}
    \,.
  \end{align}
  Explicitly evaluating the $\mathbf{P}_\perp$ integral
  gives the following result for the \emph{full} stress tensor:
\begin{align}
  T^{ij}(\mathbf{b}_\perp;\mathbf{k}_\perp,\sigma)
  &=
  \frac{ 1 }{2P^+}
  \int \frac{\mathrm{d}^2\boldsymbol{\Delta}_\perp}{(2\pi)^2}
  \Bigg\{
    \left(
    2
    \mathbf{k}_\perp^i
    \mathbf{k}_\perp^j
    +
    \frac{ \delta^{ij} }{2\sigma^2}
    \right)
    \mathcal{A}(t)
    +
    \frac{1}{2}
    \Big(
    \boldsymbol{\Delta}_\perp^i \boldsymbol{\Delta}_\perp^j
    -
    \delta^{ij} \boldsymbol{\Delta}_\perp^2
    \Big)
    \mathcal{D}(t)
    \Bigg\}
  e^{-i\boldsymbol{\Delta}_\perp\cdot\mathbf{b}_\perp}
  e^{-\frac{\sigma^2}{2} \boldsymbol{\Delta}_\perp^2}
  \,.
\end{align}
For the photon in a helicity state, $\mathcal{A}(t) = \mathcal{D}(t) = F_1(t)$.
Using this, and noting that
the radial pressure is obtained by contracting the stress tensor with
$\hat{b}_i \hat{b}_j$(with $\hat{b}$ being the unit radial vector in the transverse plane),
we find:
\begin{align}
  \label{eqn:pressure:total:0}
  \mathscr{P}_r(\mathbf{b}_\perp;\mathbf{k}_\perp,\sigma)
  =
  \frac{ 1 }{4P^+}
  \left(
  4 \mathbf{k}_\perp^2 \cos^2(\phi_{bk})
  +
  \frac{ 1 }{\sigma^2}
  +
  \frac{1}{b_\perp}
  \frac{\mathrm{d}}{\mathrm{d}b_\perp}
  \right)
  \int \frac{\mathrm{d}^2\boldsymbol{\Delta}_\perp}{(2\pi)^2}
  F_1(t)
  e^{-i\boldsymbol{\Delta}_\perp\cdot\mathbf{b}_\perp}
  e^{-\frac{\sigma^2}{2} \boldsymbol{\Delta}_\perp^2}
  \,,
\end{align}
where $\phi_{bk} = \phi - \phi_\Delta$
and a script $\mathscr{P}_r$ is used to differentiate from the intrinsic
radial pressure $p_r$.
Unlike $p_r$, $\mathscr{P}_r$ contains contributions from $\mathcal{A}(t)$,
which include both the average motion of the photon and statistical motion
due to wave function dispersion.

For the free photon, the transverse radiation pressure given by
Eq.~(\ref{eqn:pressure:total:0}) is strictly non-negative.
When $F_1(t)=1$, as in the free case, one has:
\begin{align}
  \int \frac{\mathrm{d}^2\boldsymbol{\Delta}_\perp}{(2\pi)^2}
  e^{-i\boldsymbol{\Delta}_\perp\cdot\mathbf{b}_\perp}
  e^{-\frac{\sigma^2}{2} \boldsymbol{\Delta}_\perp^2}
  =
  \frac{1}{2\pi\sigma^2} e^{-\frac{1}{2\sigma^2} \mathbf{b}_\perp^2}
  \,,
\end{align}
and it is easy to see that:
\begin{align}
  \left(
  \frac{ 1 }{\sigma^2}
  +
  \frac{1}{b_\perp}
  \frac{\mathrm{d}}{\mathrm{d}b_\perp}
  \right)
  e^{-\frac{1}{2\sigma^2} \mathbf{b}_\perp^2}
  =
  0
  \,.
\end{align}
The $\mathbf{k}_\perp^2$ term is strictly positive at $\mathbf{k}_\perp\neq0$
by virtue of positivity of the light front momentum density,
which is guaranteed by the probability interpretation of the
``good'' components of light front densities
(in this case, $T^{++}$)~\cite{Melosh:1974cu,Leutwyler:1977vy}.
When $\sigma\rightarrow0$, the total radial pressure becomes a delta function,
and the free photon accordingly behaves like a pointlike particle
that exerts a positive total pressure.

The radiation pressure described by Eq.~(\ref{eqn:pressure:total:0})
is non-negative even for the interacting photon,
and this holds to all orders in any gauge-invariant theory of photon interactions.
In fact, the only assumptions we need to prove non-negativity
are Eq.~(\ref{eqn:pressure:total:0})
and the positivity of $\rho_{p^+}(b_\perp)$.
Let us first consider the $\mathbf{k}_\perp=0$ case,
since the $\mathbf{k}_\perp^2$ term is positive anyway.
Using the convolution theorem,
Eq.~(\ref{eqn:pressure:total:0}) can be rewritten:
\begin{align}
  \int \frac{\mathrm{d}^2\boldsymbol{\Delta}_\perp}{(2\pi)^2}
  F_1(t)
  e^{-i\boldsymbol{\Delta}_\perp\cdot\mathbf{b}_\perp}
  e^{-\frac{\sigma^2}{2} \boldsymbol{\Delta}_\perp^2}
  =
  \frac{1}{2\pi \sigma^2 P^+}
  \int \mathrm{d}^2 \mathbf{b}'_\perp \,
  \rho_{p^+}(b'_\perp)
  e^{-\frac{1}{2\sigma^2}(\mathbf{b}_\perp-\mathbf{b}'_\perp)^2}
  \,,
\end{align}
where $\rho_{p^+}(b_\perp)$ was defined in
Eq.~(\ref{eqn:p+:hel}).
\deleted{of \companion.}
From Eq.~(\ref{eqn:pressure:total:0}), the radiation pressure
for $\mathbf{k}_\perp=0$ is then:
\begin{align}
  \mathscr{P}_r(\mathbf{b}_\perp;\mathbf{k}_\perp=0,\sigma)
  =
  \frac{1}{8\pi\sigma^2(P^+)^2}
  \int \mathrm{d}^2 \mathbf{b}'_\perp \,
  \rho_{p^+}(b'_\perp)
  \left(
  \frac{ 1 }{\sigma^2}
  +
  \frac{1}{b_\perp}
  \frac{\mathrm{d}}{\mathrm{d}b_\perp}
  \right)
  e^{-\frac{1}{2\sigma^2}(\mathbf{b}_\perp-\mathbf{b}'_\perp)^2}
  \,.
\end{align}
Performing the derivative with respect to $b_\perp = |\mathbf{b}_\perp|$
gives:
\begin{align}
  \mathscr{P}_r(\mathbf{b}_\perp;\mathbf{k}_\perp=0,\sigma)
  =
  \frac{1}{8\pi\sigma^4(P^+)^2}
  \int \mathrm{d}^2 \mathbf{b}'_\perp \,
  \rho_{p^+}(b'_\perp)
  \frac{\mathbf{b}_\perp\cdot\mathbf{b}'_\perp}{b_\perp^2}
  e^{-\frac{1}{2\sigma^2}(\mathbf{b}_\perp-\mathbf{b}'_\perp)^2}
  \,.
\end{align}
Using polar coordinates, with $\phi$ signifying the angle between
$\mathbf{b}_\perp$ and $\mathbf{b}'_\perp$, gives:
\begin{align}
  \mathscr{P}_r(\mathbf{b}_\perp;\mathbf{k}_\perp=0,\sigma)
  =
  \frac{1}{8\pi\sigma^4(P^+)^2 b_\perp}
  e^{-\frac{1}{2\sigma^2}b_\perp^2}
  \int \mathrm{d} b'_\perp \,
  b'^2_\perp
  \rho_{p^+}(b'_\perp)
  e^{-\frac{1}{2\sigma^2}b'^2_\perp}
  \int \mathrm{d}\phi \,
  \cos\phi
  \,
  e^{+\frac{b_\perp b'_\perp}{\sigma^2}\cos\phi}
  \,.
\end{align}
Now, the quantity
\begin{align}
  \int_0^{2\pi} \mathrm{d}\phi \,
  e^{+\frac{b_\perp b'_\perp}{\sigma^2}\cos\phi}
  \cos\phi
  =
  2\pi
  I_1\left(\frac{b_\perp b_\perp'}{\sigma^2}\right)
  \,,
\end{align}
which is a modified Bessel function of the first kind~\cite{NIST:DLMF},
is positive since it receives larger weights from the exponential
when $\cos\phi$ is positive than when $\cos\phi$ is negative.
The remaining factors in the $b'_\perp$ integral are non-negative.
If $\rho_{p^+}(b'_\perp)$ is a delta function (as in the free case),
the result is zero (as we saw explicitly in the free case),
but if $\rho_{p^+}(b'_\perp)$ is positive in any extended region,
then it follows that
$\mathscr{P}_r(\mathbf{b}_\perp;\mathbf{k}_\perp=0,\sigma)$ is positive.
Therefore, we find that the transverse radiation pressure of a photon
is strictly non-negative, and is in fact positive if the light front
momentum density has any finite spatial extent (as it does in QED).


\section{Summary and conclusions}
\label{sec:conclusion}

We calculated the momentum, angular momentum, and pressure densities of
a photon on the light front,
both for a free photon and for a QED photon at leading order.
We calculated both the intrinsic pressure and the radiation pressure,
clarifying the difference between these.
The intrinsic pressure in particular is encoded by the D-term,
and exhibits different properties than the radiation pressure,
only the latter of which is positive for a photon.
The D-term of the photon is positive,
\replaced{
in stark contrast to the negativity criterion for stability of massive systems,
}{
contradicting the hypothesis that a negative D-term is required for stability,
}
and entailing that the intrinsic pressure density of a photon is negative.

\begin{acknowledgments}
  The authors would like to thank Ian Clo\"et and Gerald Miller
  for enlightening discussions that helped contribute to this work.
  AF was supported by the U.S.\ Department of Energy
  Office of Science, Office of Nuclear Physics under Award Number
  DE-FG02-97ER-41014.
  WC was supported by the National Science Foundation under Award No.
2111442.
\end{acknowledgments}


\appendix

\section{Hankel transform of logarithm}
\label{app:polar}

To the best of our knowledge, the 0th order Hankel transform of a logarithm
has not been tabulated in the mathematics or physics literature.
We will prove in this section that:
\begin{align}
  \label{eqn:hankel:log}
  \mathscr{H}_0\Big[\log(k)\Big](b)
  =
  \int_0^\infty \mathrm{d}k \,
  k J_0(bk) \log(k)
  =
  -
  \frac{\Theta(b)}{b^2}
  \,,
\end{align}
where $\Theta(b)$ is the Heaviside step function.

To obtain this result,
first consider the variable transformation $k \mapsto sk$
in the original integral:
\begin{align}
  F(b)
  \equiv
  \mathscr{H}_0\Big[\log(k)\Big](b)
  =
  s^2
  \int_0^\infty \mathrm{d}k \,
  k J_0(bsk) \log(sk)
  =
  s^2
  F(sb)
  +
  s^2 \log(s)
  \int_0^\infty \mathrm{d}k \,
  k J_0(bsk)
  \,.
\end{align}
Next, note that
the Hankel transform of a constant is a delta function:
\begin{align}
  \int_0^\infty \mathrm{d}k \,
  k J_0(bk)
  =
  \frac{\delta(b)}{b}
  \,.
\end{align}
This gives us:
\begin{align}
  F(sb)
  =
  \frac{1}{s^2}
  F(b)
  -
  \log(s)
  \frac{\delta(b)}{b}
  \,.
\end{align}
Differentiating with respect to $s$
and then
taking $s=1$ gives a differential equation for $F(b)$:
\begin{align}
  b F'(b) + 2 F(b)
  =
  -2\pi
  \frac{\delta(b)}{b}
  \,.
\end{align}
This can be solved by substituting the ansatz:
\begin{align}
  F(b)
  =
  \frac{f(b)}{b^2}
  \,,
\end{align}
with $f(b)$ satisfying:
\begin{align}
  \frac{1}{b} f'(b)
  =
  -
  \frac{\delta(b)}{b}
  \,.
\end{align}
This has the solution:
\begin{align}
  f(b)
  =
  C - \Theta(b)
  \,,
\end{align}
and therefore the desired Hankel transform is:
\begin{align}
  F(b)
  =
  \frac{C-\Theta(b)}{b^2}
  \,.
\end{align}

To determine $C$, we can find the mean-squared radius associated with $F(b)$
when it has been multiplied by a Gaussian falloff.
Unless $C=1$, the radius associated with $F(b)$ itself will diverge,
so the Gaussian falloff factor is necessary.
The quantity in question is defined:
\begin{align}
  \langle b^2 \rangle(\sigma)
  \equiv
  \int \mathrm{d}^2\mathbf{b} \,
  b^2
  F(b)
  e^{-\frac{b^2}{2\sigma^2}}
  \,.
\end{align}
From direct evaluation, we find:
\begin{align}
  \langle b^2 \rangle(\sigma)
  =
  \sigma^2(C - 1)
  \,.
\end{align}
Now, we compare to the radius result obtained by evaluating in momentum space.
Using the convolution theorem, the product in this integral
can be written as the 2D Fourier transform of a convolution:
\begin{align}
  F(b)
  e^{-\frac{b^2}{2\sigma^2}}
  =
  \int \frac{\mathrm{d}^2\mathbf{k}}{(2\pi)^2}
  e^{-i\mathbf{k}\cdot\mathbf{b}}
  \int \frac{\mathrm{d}^2\mathbf{k}'}{(2\pi)^2}
  \left( 2\pi \sigma^2 e^{-\frac{\sigma^2}{2} (\mathbf{k}-\mathbf{k}')^2} \right)
  \log(k')
  \,.
\end{align}
Using integration by parts, the mean-squared radius can be written:
\begin{align}
  \langle b^2 \rangle(\sigma)
  &=
  -
  2\pi \sigma^2
  \int \frac{\mathrm{d}^2\mathbf{k}'}{(2\pi)^2}
  \nabla_{\mathbf{k}}^2 \left[
    e^{-\frac{\sigma^2}{2} (\mathbf{k}-\mathbf{k}')^2}
    \right]
  \Bigg|_{\mathbf{k}=0}
  \log(k')
  =
  \frac{\sigma^4}{2}
  \int_0^\infty \mathrm{d}\kappa\,
  \left( 1 - \frac{1}{2} \sigma^2 \kappa \right)
  e^{-\frac{\sigma^2}{2}\kappa}
  \log\kappa
  \,.
\end{align}
Evaluating this integral gives:
\begin{align}
  \langle b^2 \rangle(\sigma)
  =
  -\sigma^2
  \,,
\end{align}
thus requiring $C=0$, and proving Eq.~(\ref{eqn:hankel:log}).


\bibliography{references.bib}

\end{document}